\documentclass[letterpaper,twocolumn,10pt]{article}

\usepackage{zhanggroup}
\usepackage{tikz}
\usepackage{xspace} 
\usepackage{mathrsfs}
\usepackage{amsfonts}
\usepackage{amsthm}
\usepackage{subcaption}
\usepackage{booktabs}
\usepackage{multirow}
\usepackage{diagbox}
\usepackage{graphicx}
\usepackage{url}
\usepackage{caption,subcaption}
\usepackage{amsmath}
\usepackage{dsfont}
\usepackage{mathtools}
\usepackage{colortbl}
\usepackage{multirow}
\usepackage{array}
\usepackage{makecell}
\usepackage{algorithm2e}
\usepackage[absolute]{textpos}
\usepackage{titlesec}
\SetKwComment{Comment}{/* }{ */}
\RestyleAlgo{ruled}
\hypersetup{
  colorlinks,
  linkcolor={blue!70!green},
  citecolor={green!70!blue},
  urlcolor={orange!70!red}
}

\newcommand{\mypara}[1]{\smallskip\noindent{\bf {#1}.}}
\newcommand{\SSLGuard}{\emph{SSLGuard}\xspace}
\newcommand{\stealone}{\texttt{Steal-1}\xspace}
\newcommand{\stealtwo}{\texttt{Steal-2}\xspace}
\newcommand{\stealthree}{\texttt{Steal-3}\xspace}

\begin{document}

\begin{textblock}{15}(1.9,1)
To Appear in 2022 ACM SIGSAC Conference on Computer and Communications Security, November 2022
\end{textblock}

\title{\Large \bf SSLGuard: A Watermarking Scheme for Self-supervised Learning \\ Pre-trained Encoders}

\author{
Tianshuo Cong\textsuperscript{1}\ \ \
Xinlei He\textsuperscript{2}\ \ \
Yang Zhang\textsuperscript{2}
\\
\\
\textsuperscript{1}\textit{Institute for Advanced Study, BNRist, Tsinghua University}\ \ \ \\
\textsuperscript{2}\textit{CISPA Helmholtz Center for Information Security}
}

\date{}

\maketitle

\begin{abstract}

Self-supervised learning is an emerging machine learning paradigm. 
Compared to supervised learning which leverages high-quality labeled datasets, self-supervised learning relies on unlabeled datasets to pre-train powerful encoders which can then be treated as feature extractors for various downstream tasks.
The huge amount of data and computational resources consumption makes the encoders themselves become the valuable intellectual property of the model owner.
Recent research has shown that the machine learning model's copyright is threatened by \emph{model stealing attacks}, which aim to train a surrogate model to mimic the behavior of a given model.
We empirically show that pre-trained encoders are highly vulnerable to model stealing attacks.
However, most of the current efforts of copyright protection algorithms such as watermarking concentrate on classifiers.
Meanwhile, the intrinsic challenges of pre-trained encoder's copyright protection remain largely unstudied.
We fill the gap by proposing \SSLGuard, the first watermarking scheme for pre-trained encoders.
Given a clean pre-trained encoder, \SSLGuard injects a watermark into it and outputs a watermarked version.
The shadow training technique is also applied to preserve the watermark under potential model stealing attacks.
Our extensive evaluation shows that \SSLGuard is effective in watermark injection and verification, and it is robust against model stealing and other watermark removal attacks such as input noising, output perturbing, overwriting, model pruning, and fine-tuning.\footnote{Our code is available at \url{https://github.com/tianshuocong/SSLGuard}.}

\end{abstract}

\section{Introduction}
\label{section:introduction}

Deep learning, in particular supervised learning (SL), has gained tremendous success during the past decade, and the development of SL relies on a large amount of high-quality labeled data.
However, high-quality data is often difficult to collect and the cost of labeling is expensive.
\emph{Self-supervised learning (SSL)} is proposed to resolve such restrictions by generating ``labels'' from the unlabeled dataset (called \emph{pre-training dataset}) and uses the derived ``labels'' to pre-train an \emph{encoder} which can output informative embeddings.
SSL encoders have shown great promise in various downstream tasks.
For instance, on the ImageNet dataset~\cite{RDSKSMHKKBBF15}, Chen et al.~\cite{CKNH20} show that, by using SimCLR pre-trained with ImageNet (unlabeled), the downstream classifier can achieve $85.8\%$ top-5 accuracy with only $1\%$ labels, which outperforms a supervised AlexNet but uses 100$\times$ fewer labels.
He et al.~\cite{HFWXG20} show that SSL can surpass SL under 7 downstream tasks including segmentation and detection.
Therefore, compared to the SL-based classifier
which only suits a specific classification task, the SSL pre-trained encoder can achieve remarkable performance on different downstream tasks.

\begin{figure}[t]
\centering
\includegraphics[width=6cm]{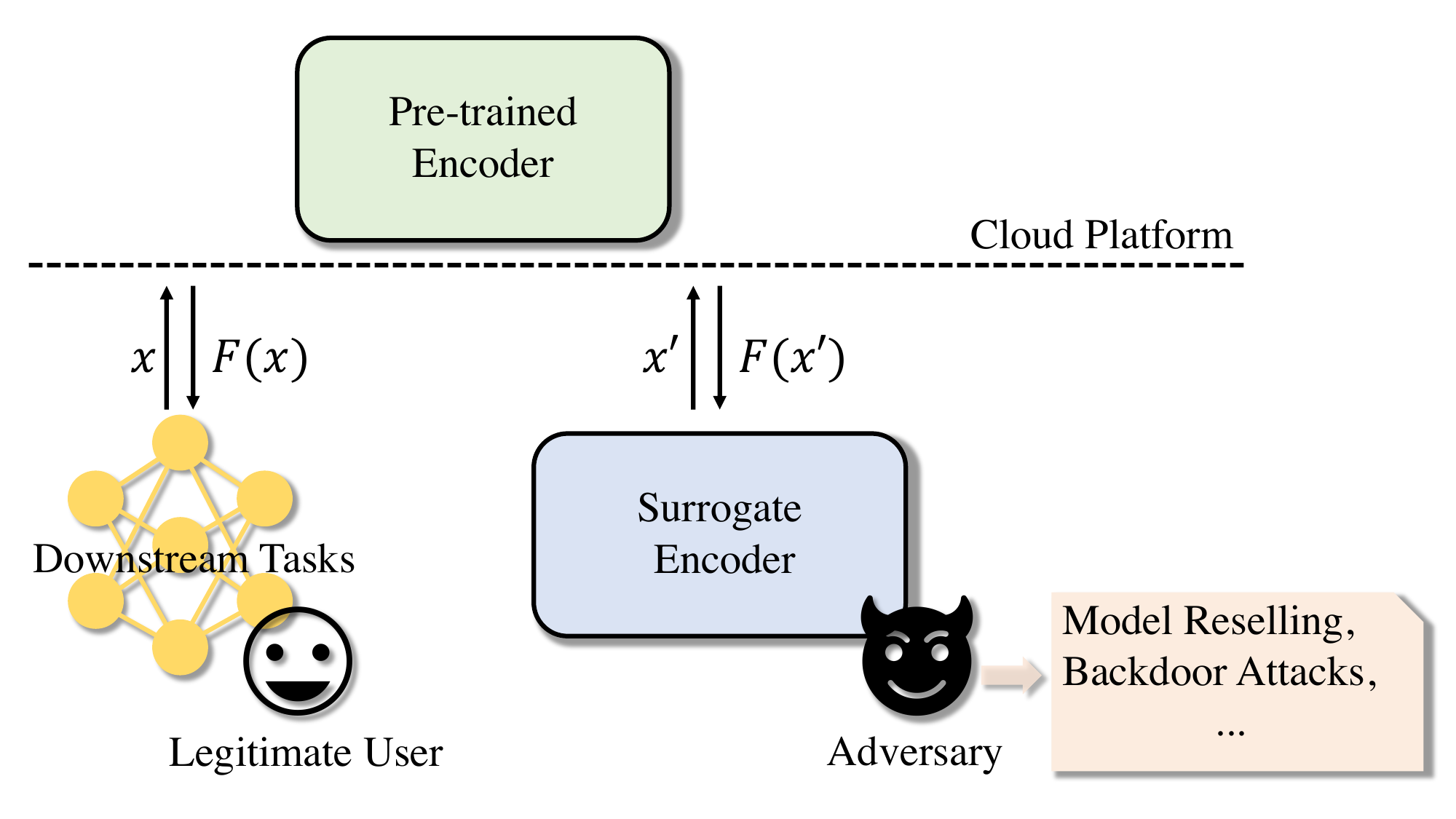}
\caption{An illustration of deploying an SSL pre-trained encoder as a service. The legitimate user aims to train downstream classifiers while the adversary tries to generate a surrogate encoder.}
\label{fig::scenario}
\end{figure}

However, the data collection and training process of SSL encoders are also expensive as they benefit from larger datasets and more powerful computing devices.
For example, the performance of MoCo~\cite{HFWXG20} pre-trained with the Instagram-1B dataset ($\sim1$ billion images) outperforms that of the encoder pre-trained with the ImageNet-1M dataset (1.28 million images), and SimCLR requires 32 TPU v3 cores to train a ResNet-50 due to the large batch size setting (i.e., 4096)~\cite{CKNH20}.
Therefore, the cost to train a powerful encoder by SSL is prohibitive for individuals, and the high-performance encoders are usually pre-trained by leading AI companies with sufficient computing resources and shared via cloud platforms for commercial usage, i.e., Encoder-as-a-Service (EaaS)~\cite{OpenAIAPI, ClarifaiAPI}.
For instance, Clarifai~\cite{ClarifaiAPI} provides image encoders for different downstream services.
OpenAI provides access to GPT-3~\cite{BMRSKDNSSAAHKHCRZWWHCSLGCCBMRSA20} which can be considered as a powerful encoder for a variety of natural language processing (NLP) downstream tasks, such as code generation, style transfer, etc.

Once deployed on the cloud platform, the encoders are not only accessible to legitimate users but also threatened by potential adversaries.
As illustrated in \autoref{fig::scenario}, for the legitimate user, the encoder is used to train a downstream classifier.
On the other hand, an adversary may perform \emph{model stealing attacks}~\cite{TZJRR16, OSF19, KTPPI20, SHHZ22} which aim to learn a surrogate encoder that has similar functionality.
Such attacks may not only compromise the intellectual property of the service provider but also serve as a stepping stone for further attacks such as membership inference attacks (MIA)~\cite{SSSS17, SZHBFB19, LJQG21} (i.e., mount MIA offline by using surrogate encoders), backdoor attacks~\cite{JLG22} (i.e., publish another backdoored encoder), and adversarial attacks~\cite{PMGJCS17}.
The security and privacy of SSL encoders are threatened by these attacks, which call for effective defenses.

As one major technique to protect the machine learning model's copyright, model watermarking~\cite{LHZG19, JCCP21} inserts a secret pattern into the model.
Then, the ownership can be claimed if a similar or the same pattern is successfully extracted from the model.
Recent studies on model watermarking mainly focus on the classifier that targeted a specific task~\cite{ABCPK18, ZGJWSHM18, JCCP21}.
However, watermarking SSL encoders may face several intrinsic challenges.
First, model watermarking against the classifier usually needs to specify a target class, while the SSL encoder does not have such information.
Second, downstream tasks for SSL encoders are flexible, which challenges the traditional model watermarking scheme that is only suitable for one specific downstream task.
Therefore, a new watermarking scheme should be designed to overcome those challenges to protect the copyright of SSL encoders.
To the best of our knowledge, this has been left largely unstudied.

\mypara{Our Work}
In this paper, we first quantify the copyright breaching threat against SSL encoders through the lens of model stealing attacks.
Then, we introduce \SSLGuard, the first watermarking scheme for the SSL encoders to protect their copyrights.
Note that in this work, we consider image encoders only.

For model stealing attacks, we first assume that the adversary only has black-box access to the victim encoder.
We then characterize the adversary's background knowledge into two dimensions, i.e., the surrogate dataset and the surrogate encoder's architecture.
Regarding the surrogate dataset which is used to train the surrogate encoder, we consider the adversary may or may not know the victim encoder's pre-training dataset.
Regarding the surrogate encoder's architecture, we first assume that it shares the same architecture as the victim encoder.
Then, we relax this assumption and find that the effectiveness of model stealing attacks can even increase by leveraging a larger model architecture.
We empirically show that the model stealing attacks achieve remarkable performance.
For instance, given a ResNet-50 encoder pre-trained on ImageNet by SimCLR, the ResNet-101 surrogate encoder can achieve 0.944 accuracy on STL-10 while the accuracy for the victim encoder is 0.948.
We also show that the cost of stealing an encoder is much smaller than pre-training it from scratch, e.g., pre-training a BYOL ResNet-50 encoder costs \$5,713.92 while stealing it with ResNet-101 only costs \$72.49 (see \autoref{table:money} for the detailed comparison).
Such observation emphasizes the underlying threat of jeopardizing the model owner's intellectual property and the emergence of copyright protection.

To protect the copyright of SSL encoders, we propose a robust \emph{black-box} watermarking scheme named \SSLGuard.
Concretely, the goal of \SSLGuard is to inject a watermark based on a given secret vector into a clean SSL encoder.
The output of \SSLGuard contains a watermarked encoder and a key-tuple.
To be specific, the key-tuple consists of the secret vector, a verification dataset, and a decoder.
\SSLGuard fine-tunes a clean encoder to a watermarked encoder which can keep the utility and map samples in the verification dataset to secret embeddings.
We further introduce a decoder to transform these secret embeddings into the secret vector.
For other encoders, the decoder only transforms the embeddings generated from the verification dataset into random vectors.
Recent research has shown that if a watermarked model is stolen, its corresponding watermark usually vanishes~\cite{LJLK22}.
To remedy this situation, \SSLGuard adopts a shadow dataset and a shadow encoder to locally simulate model stealing attacks.
Meanwhile, \SSLGuard optimizes a trigger that can be recognized by both the watermarked encoder and the shadow encoder.
We later show in \autoref{section:evaluation} that such a design can strongly preserve the watermark even in the surrogate encoder stolen by the adversary.

Empirical evaluations over 7 datasets (i.e., ImageNet, CIFAR-10, CIFAR-100, STL-10, GTSRB, MNIST, and FashionMNIST) and 3 encoder pre-training algorithms (i.e., SimCLR, MoCo v2, and BYOL) show that \SSLGuard can successfully inject/extract the watermark to/from the SSL encoder without sacrificing its performance and is robust to model stealing attacks.
Moreover, we consider various types of watermark removal attacks including input preprocessing (noising), output perturbing (noising and truncation), and model modification (overwriting, pruning, and fine-tuning).
We empirically show that \SSLGuard is still effective in such a scenario.

In summary, we make the following contributions:
\begin{itemize}
    \item We unveil that the SSL pre-trained encoders are highly vulnerable to model stealing attacks.
    \item We propose \SSLGuard, the first watermarking scheme against SSL pre-trained encoders, which can protect the intellectual property of published encoders.
    \item Extensive evaluations show that \SSLGuard is effective in injecting and extracting watermarks, and it is robust against model stealing and other watermark removal attacks such as input noising, output perturbing, overwriting, model pruning, and fine-tuning.
\end{itemize}

\section{Background}
\label{section:background}

\subsection{Self-supervised Learning}

Self-supervised learning is a rising AI paradigm that aims to train an encoder by a large scale of unlabeled data.
A high-performance pre-trained encoder can be shared into the public platform as an upstream service.
In downstream tasks, customers can use the embeddings output from the pre-trained encoder to train their classifiers with limited labeled data~\cite{CKNH20} or even no data~\cite{RKHRGASAMCKS21}.
One of the most remarkable self-supervised learning paradigms is contrastive learning~\cite{CKNH20,HFWXG20,CFGH20,GSATRBDPGAPKMV20,RKHRGASAMCKS21}.
In general, encoders are pre-trained through contrastive losses which calculate the similarities of embeddings in a latent space.
In this paper, we consider three representative contrastive learning algorithms, i.e., SimCLR~\cite{CKNH20}, MoCo v2~\cite{CFGH20}, and BYOL~\cite{RKHRGASAMCKS21}.

\mypara{SimCLR~\cite{CKNH20}}
SimCLR is a simple framework for contrastive learning.
It consists of 4 components, including \emph{Data augmentation}, \emph{Base encoder $f(\cdot)$}, \emph{Projection head $g(\cdot)$} and \emph{Contrastive loss function}.

The data augmentation module is used to transform a data sample $x$ randomly into two augmented views.
Specifically, the augmentations include random cropping, random color distortions, and random Gaussian blur.
If two augmented views are generated from the same data sample $x$, we treat them as a positive pair, otherwise, they are considered a negative pair.
Positive pairs of $x$ are denoted as $\tilde{x_i}$ and $\tilde{x_j}$.

Base encoder $f(\cdot)$ extracts feature vectors $h_i=f(\tilde{x_i})$ from augmented inputs.
Projection head $g(\cdot)$ is a small neural network that maps feature vectors to a latent space where contrastive loss is applied.
SimCLR uses a multilayer perceptron (MLP) as the projection head $g(\cdot)$ to obtain the output $z_i = g(h_i)$.

For a set of samples $\{\tilde{x_k}\}$ including both positive and negative pairs, contrastive loss aims to maximize the similarity between the feature vectors of positive pairs and minimize those of negative pairs.
Given $N$ samples in each mini-batch, we could get $2N$ augmented samples.
Formally, the loss function for a positive pair $\tilde{x_i}$ and $\tilde{x_j}$ can be formulated as:
\begin{equation*}
l(i,j) = -\log \frac{\exp(\text{sim}(z_i,z_j)/\tau)}{\sum^{2N}_{k=1,k\neq i}\exp(\text{sim}(z_i,z_k)/\tau)},
\end{equation*}
where $\text{sim}(\cdot,\cdot)$ denotes the cosine similarity function and $\tau$ denotes a temperature parameter.
SimCLR jointly trains the base encoder and projection head by minimizing the final loss function:
\begin{equation*}
\mathcal{L}_{SimCLR} = \frac{1}{2N}\sum_{k=1}^N[l(2k-1,2k)+l(2k,2k-1)],
\end{equation*}
where $2k-1$ and $2k$ are the indexes for each positive pair.
Once the model is trained, SimCLR discards the projection head and keeps the base encoder $f(\cdot)$ only, which serves as the pre-trained encoder.

\mypara{MoCo v2~\cite{CFGH20}}
Momentum Contrast (MoCo)~\cite{HFWXG20} is a famous contrastive learning algorithm, and MoCo v2 is the modified version (using a projection head and more data augmentations).

MoCo points out that contrastive learning can be regarded as a dictionary lookup task.
The ``keys'' in the dictionary are the embeddings output from the encoder.
A ``query'' matches a key if they are encoded from the same image.
MoCo aims to train an encoder that outputs similar embeddings for a query and its matching key, and dissimilar embeddings for others.
The dictionary is desirable to be large and consistent, which contains rich negative images and helps to learn good embeddings.
MoCo aims to build such a dictionary with a queue and momentum encoder.

MoCo contains two parts: \emph{query encoder} $f_q(x;\theta_q)$ and \emph{key encoder} $f_k(x;\theta_k)$.
Given a query sample $x^q$, MoCo gets an encoded query $q=f_q(x^q)$.
For other samples $x^k$, MoCo builds a dictionary whose keys are $\{k_0,k_1,...\}$, $k_i=f_k(x_i^k)$.
The dictionary is a dynamic queue that keeps the current mini-batch encoded embeddings and discards the ones in the oldest mini-batch.
The benefit of using a queue is decoupling the dictionary size from the mini-batch size, so the dictionary size can be set as a hyper-parameter.
Assume $k_+$ is the key that $q$ matches, the loss function will be defined as:
\begin{equation*}
\mathcal{L}_{MoCo} = -\log\frac{\exp(q\cdot k_+/\tau)}{\sum_{i=0}^K \exp(q\cdot k_i/\tau)}.
\end{equation*}
Here $\tau$ is a temperature hyper-parameter.
MoCo trains $f_q$ by minimizing contrastive loss and updates $\theta_q$ by gradient descent.
However, it is difficult to update $\theta_k$ by back-propagation because of the queue, so $f_k$ is updated by moving-averaged as:
\begin{equation*}
\theta_k \leftarrow m \theta_k + (1-m)\theta_q,
\end{equation*}
where $m \in [0,1)$ denotes a momentum coefficient.
Finally, we keep the $f_q$ as the final pre-trained encoder.

\mypara{BYOL~\cite{GSATRBDPGAPKMV20}}
Bootstrap Your Own Latent (BYOL) is a novel self-supervised learning algorithm.
Different from previous methods, BYOL does not rely on negative pairs, and it has a more robust selection of image augmentations.

BYOL's architecture consists of two neural networks: \emph{online networks} and \emph{target networks}.
The online networks, with parameters $\theta$, consist of an encoder $f_{\theta}$, a projector $g_{\theta}$ and a predictor $q_{\theta}$.
The target networks are made up of an encoder $f_{\xi}$ and a projector $g_{\xi}$.
The two networks bootstrap the embeddings and learn from each other.

Given an input sample $x$, BYOL produces two augmented views $v \leftarrow t(x)$ and $v' \leftarrow t'(x)$ by using image augmentations $t$ and $t'$, respectively.
The online networks output a projection $z_{\theta} \leftarrow g_{\theta}(f_{\theta}(v))$ and target networks output 
a target projection $z'_{\xi} \leftarrow g_{\xi}(f_{\xi}(v'))$.
The online networks' goal is to make the prediction $q_\theta(z_\theta)$ similar to $z'_{\xi}$.
Formally, the similarity can be defined as the following:

\begin{equation*}
L_{\theta,\xi} = 2 - 2 \cdot \frac{\langle q_{\theta}(z_{\theta}), z'_{\xi} \rangle}{\Vert q_{\theta}(z_{\theta}) \Vert_2 \cdot \Vert z'_{\xi} \Vert_2}.
\end{equation*}

Conversely, BYOL feeds $v'$ to the online networks and $v$ to the target networks separately and gets $ \widetilde{L_{\theta,\xi}}$.
The final loss function can be formulated as:

\begin{equation*}
\mathcal{L}_{BYOL} = L_{\theta,\xi} + \widetilde{L_{\theta,\xi}}.
\end{equation*}

BYOL updates the weights of the online and target networks by:
\begin{equation*}
\theta \leftarrow {\rm optimizer}(\theta, \bigtriangledown_{\theta}L_{\theta,\xi}^{BYOL},\eta),
\end{equation*}
\begin{equation*}
\xi \leftarrow \tau \xi + (1-\tau)\theta,
\end{equation*}
where $\eta$ is the learning rate of the online networks.
The target networks' weight $\xi$ is updated in a weighted average way, and $\tau \in [0,1]$ denotes the decay rate of the target encoder.
Once the model is trained, we treat the online networks' encoder $f_{\theta}$ as the pre-trained encoder.

\subsection{Model Stealing Attacks}

Model stealing attacks~\cite{TZJRR16,CCGJY18,OSF19,DR19,KTPPI20,JCBKP20,CCGJY20,WYPY20,SHHZ22} aim to steal the parameters or the functionality of the victim model.
To achieve this goal, given a victim model $f(x;\theta)$, the adversary can issue a bunch of queries to the victim model and obtain the corresponding responses.
Then the queries and responses serve as the inputs and ``labels'' to train the surrogate model, denoted as $f'(x;\theta')$.
Formally, given a query dataset $\mathcal{D}$, the adversary can train $f'(x;\theta')$ by
\begin{equation}
\label{equ:modelstealing}
\mathcal{L}_{steal} = \mathbb{E}_{x \sim \mathcal{D}}[{\rm sim}(f(x;\theta),f'(x;\theta'))],
\end{equation}
where ${\rm sim(\cdot,\cdot)}$ is a similarity function.

Note that if the victim model is a classifier, the response can be the prediction probability of each class.
If the victim model is an encoder, the response can be the embeddings.
A successful model stealing attack may not only breach the intellectual property of the victim model but also serve as a springboard for further attacks such as MIA~\cite{LLHYBZ222,HLXCZ22,SSSS17,SZHBFB19,LJQG21,HZ21,SM21,LZ21,HWWBSZ21,HLGZ22}, backdoor attacks~\cite{YLZZ19,SSP20,CSBMSWZ21,JLG22} and adversarial attacks~\cite{GSS15,PMGJCS17,CW172,KGB16,LCLS16}.
Previous work has demonstrated that neural networks are vulnerable to model stealing attacks.
In this paper, we concentrate on model stealing attacks on SSL encoders, which have not been studied yet.

\subsection{DNNs Watermarking}

Considering the cost of training deep neural networks (DNNs), DNNs watermarking algorithms have received wide attention as it is an effective method to protect the copyright of the DNNs.
Watermarking is a traditional concept for media such as audio and video, and it has been extended to protect the intellectual property of machine learning models recently~\cite{UNSS17,MPT17,RCK18,ABCPK18,JCCP21}.
Concretely, the watermarking procedure can be divided into two steps, i.e., injection and verification.
In the injection step, the model owner injects a watermark and a pre-defined behavior into the model in the training process.
The watermark is usually secret, such as a trigger that is only known to the model owner~\cite{LHZG19}.
In the verification step, the ownership of a suspect model can be claimed if the watermarked encoder has the pre-defined behavior when the input samples contain the trigger.

So far, the watermarking algorithms mainly focus on the classifiers in a specific task.
However, how to design a watermarking algorithm for SSL pre-trained encoders that can fit various downstream tasks remains largely unexplored.

\section{Threat Model}

In this paper, we consider two parties: the \emph{defender} and the \emph{adversary}.
The defender is the owner of the victim encoder, whose goal is to protect the copyright of the victim encoder when publishing it as an online service.
The adversary, on the contrary, aims to steal the victim encoder, i.e., by model stealing attacks or directly obtaining the model (insider threat), and bypass the copyright protection method for the victim encoder.

\mypara{Adversary's Motivation}
Adversary's motivation lies in two areas: 
Firstly, EaaS is being popular and high-performance SSL encoders are often pre-trained by top AI companies~\cite{ClarifaiAPI,OpenAIAPI}.
Pre-training an encoder requires collecting a huge amount of data, expert knowledge for designing architectures/algorithms, and many failure trials, which are expensive.
This makes the model architectures or training algorithms regarded as trade secrets and will not be publicly available, which makes it less possible for the adversary to directly train a comparable performance SSL encoder from scratch.
Secondly, the cost of stealing an SSL encoder is quite less than training an SSL encoder from scratch.
For instance, pre-training a ResNet-50 by BYOL needs \$5,713.92 while generating a surrogate encoder with similar performance only needs \$72.49 (see \autoref{table:money} for more details).
Once the adversary steals the victim encoder successfully, they can resell it or deploy it on the cloud platform to be a commercial competitor.

\mypara{Adversary's Background Knowledge}
For the adversary, we first assume that they only have black-box access to the victim encoder, which is the most challenging setting for the model stealing attacks~\cite{OSF19,JCBKP20,KTPPI20,SHHZ22}.
In this setting, the adversary can only query the victim encoder with data samples and obtain their corresponding responses, i.e., the embeddings, to train the surrogate encoders.
We categorize the adversary's background knowledge into two dimensions, i.e., the pre-training dataset and the victim encoder's architecture.
Concretely, we assume that the adversary has a query dataset to perform the attack.
Note that the query dataset does not need to be in the same distribution as the victim encoder's pre-training dataset.
Regarding the victim encoder's architecture, we first assume that the adversary can obtain it since such information is usually publicly accessible.
Then we empirically show that this assumption can be relaxed, and the attack is even more effective when the adversary leverages a deeper model architecture.

\mypara{Adaptive Adversary}
We then consider an adaptive adversary who knows that the victim encoder has already been watermarked.
This means they can leverage watermark removal techniques including input preprocessing (noising), output perturbing (noising and truncation), and model modification (overwriting, pruning, and fine-tuning) on the encoder to bypass the watermark verification.

\section{Design of Watermarking Scheme}
\label{section:watermark_design}

In this section, we present \SSLGuard, a watermarking scheme to preserve the copyright of the SSL pre-trained encoders.
\SSLGuard should have the following properties:

\begin{itemize}
    \item {\bf Fidelity:} To minimize the impact of \SSLGuard on the legitimate users, the influence of \SSLGuard on the clean pre-trained encoders should be negligible, which means \SSLGuard should keep the utility of downstream tasks.
    \item {\bf Effectiveness:} \SSLGuard should judge whether a suspect model is a watermarked (or a clean) model with high precision. In other words, \SSLGuard should extract watermarks from watermarked encoders effectively.
    \item {\bf Undetectability:} The watermark cannot be extracted by a \emph{no-matching} secret key-tuple. Undetectability ensures that ownership of the SSL pre-trained encoder could not be misrepresented.
    \item {\bf Efficiency:} \SSLGuard should inject and extract watermark efficiently. For instance, the time cost for the watermark injection and extraction process should be less than pre-training an SSL model.
    \item {\bf Robustness:} \SSLGuard should be robust against model stealing attacks and other watermark removal attacks such as input noising, output perturbing, overwriting, model pruning, and fine-tuning.
\end{itemize}

In the following subsections, we will introduce the design methods for \SSLGuard.
\autoref{table:notation} summarizes the notations used in this paper.

\subsection{Overview}
\label{subsection:sslguard_overview}

\begin{figure}[t]
\centering
\includegraphics[width=8.3cm]{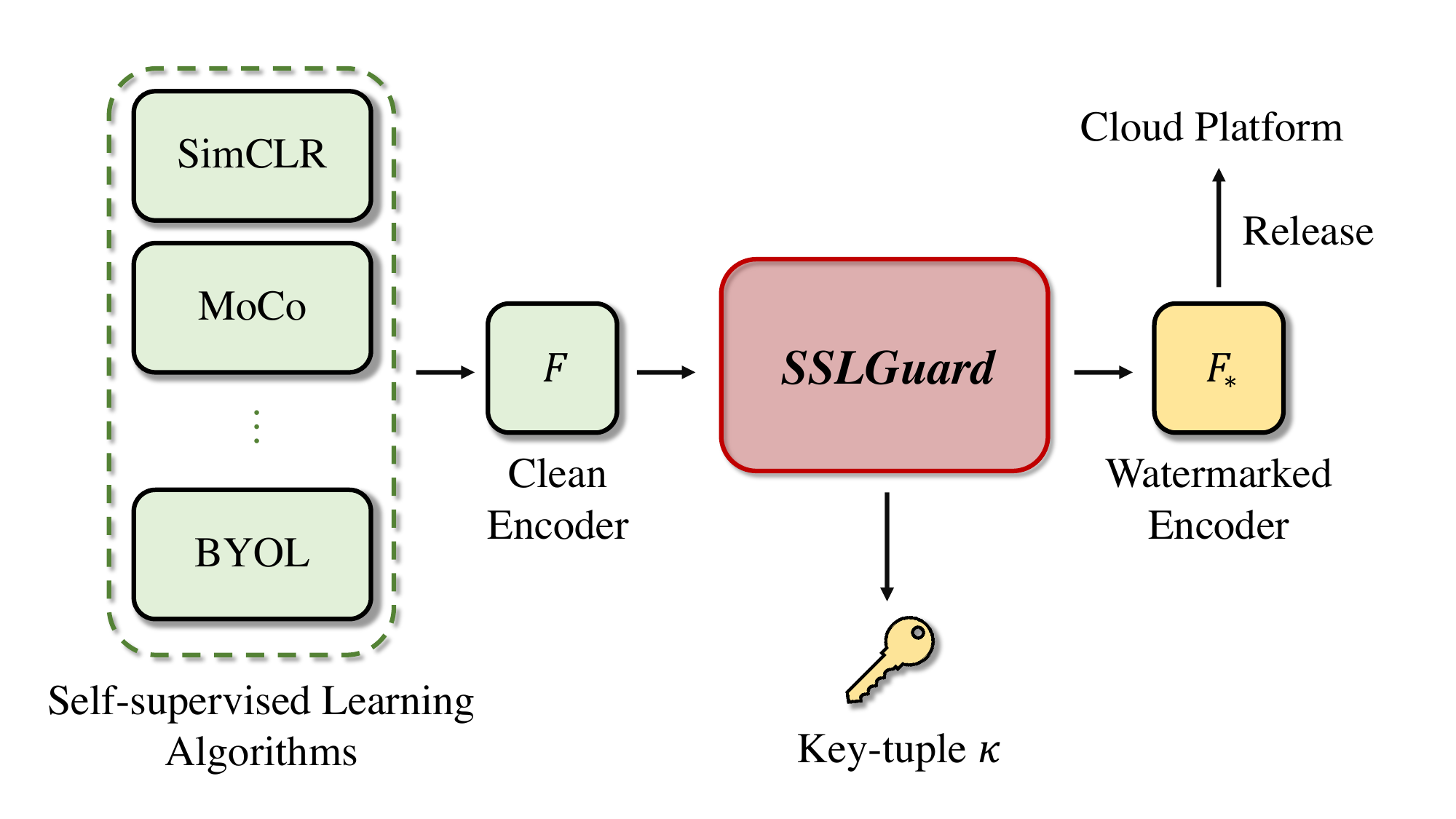}
\caption{The workflow of \SSLGuard. Given a clean SSL pre-trained encoder (colored in green), \SSLGuard outputs a key-tuple and a watermarked encoder (colored in yellow). 
The defender can employ the watermarked encoder on the cloud platform or adopt the key-tuple to extract the watermark from a suspect encoder.}
\label{fig::workflow}
\end{figure}

As shown by Cai et al.~\cite{CFJ13}, in space $\mathbb{R}^n$, given two random vectors which are independently chosen with the uniform distribution on the unit sphere, the empirical distribution of angles $\theta$ between these two random vectors converges to a distribution with the following probability density function:
\begin{equation*}
f(\theta) = \frac{1}{\sqrt \pi}\cdot \frac{\Gamma (\frac{n}{2})}{\Gamma (\frac{n-1}{2})} \cdot (\sin \theta)^{n-2}, \theta \in [0,\pi].
\end{equation*}

The distribution $f(\theta)$ will be very close to normal distribution if $n \geq 5$.
The equation above implies that two random vectors in high-dimensional space (such as $\mathbb{R}^{256}$) are almost \emph{orthogonal}.
The inspiration for \SSLGuard is based on the above mathematical fact: Given a vector that has the same dimension as embeddings, if the vector is randomly initialized, the average cosine similarity between these embeddings and the vector should be concentrated around 0.
However, if the average cosine similarity is much bigger than $0$ or even close to $1$, this can be considered as a signal that those embeddings are strongly related to this vector.
Therefore, the defender can generate a verification dataset $\mathcal{D}_{v}$ and a secret vector $sk \in \mathbb{R}^m$.
Then, the defender can fine-tune a clean encoder to transform samples from $\mathcal{D}_{v}$ to the embeddings and train a decoder to further transform the embeddings to the decoded vectors that have high cosine similarity with $sk$.
Meanwhile, if the defender input these verification samples to a clean encoder, the distribution of cosine similarity between decoded vectors and $sk$ should be a normal distribution with $0$ as its mean value.
We leverage this mechanism to design \SSLGuard.

The workflow of \SSLGuard is shown in \autoref{fig::workflow}.
Concretely, given a clean encoder $F$ which is pre-trained by a certain SSL algorithm, \SSLGuard will output a watermarked encoder $F_*$ and a secret key-tuple $\kappa$ as: 
\begin{equation*}
\begin{split}
    F_*, \kappa & \leftarrow SSLGuard(F), \\ 
    \kappa &= \{\mathcal{D}_v, G, sk\}.
\end{split}
\end{equation*}

The secret key-tuple $\kappa$ consists of three items: a verification dataset $\mathcal{D}_v$, a decoder $G$, and a secret vector $sk$.
$G$ is an MLP that maps the embeddings generated from the encoder to a new latent space (same dimension as $sk$) to calculate the cosine similarity with $sk$.
Concretely, given an input image $x$, the decoded vector $sk'_x$ can be defined as: $$sk'_x = G(E(x)), x\in \mathcal{D},$$
where $sk'_x \in \mathbb{R}^m$ is a vector whose dimension is the same as the secret vector $sk$, $\mathcal{D}$ is a given dataset, and $E$ is an encoder (i.e., $F$ or $F_*$, etc).

\begin{figure}[ht]
\centering
\includegraphics[width=7.8 cm]{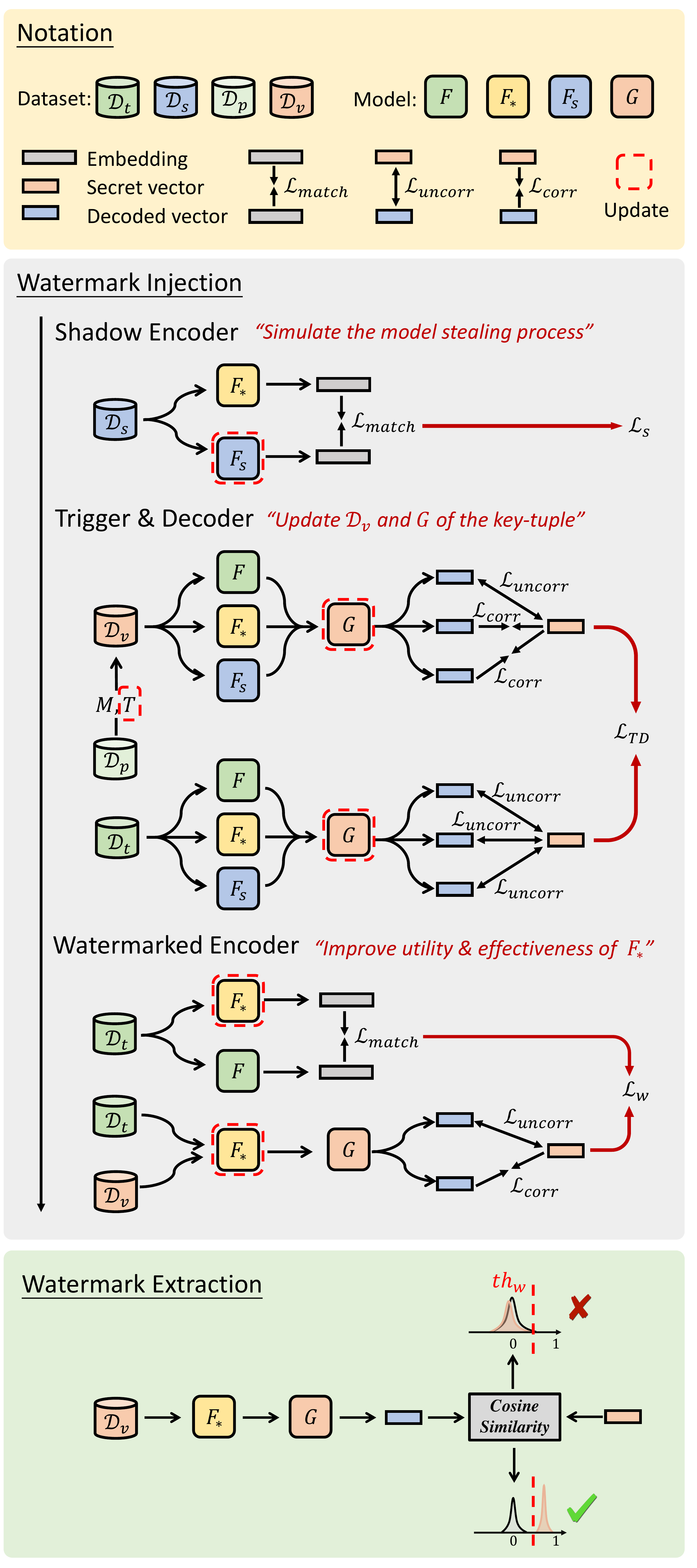}
\caption{The overview of \SSLGuard.}
\label{figure:overview}
\end{figure}

\SSLGuard contains two processes, i.e., watermark injection and extraction.
For the injection process, \SSLGuard uses a secret key-tuple $\kappa$ to inject the watermark into a clean encoder $F$ and outputs watermarked encoder $F_*$ as:
$F_* \leftarrow \texttt{Inject}(F, \kappa).$
The defender can release $F_*$ to the cloud platform and keep $\kappa$ secret.
For the extraction process, given a suspect encoder $F'$, the defender can use $\kappa$ to extract decoded vectors $sk_x'$ from $F'$ by:
$\{sk'_x\} \leftarrow \texttt{Extract}(F', \kappa), x \in \mathcal{D}_v,$
where $\{sk'_x\}$ is a set of decoded vectors.
Then, the defender can measure the cosine similarity between $\{sk'_x\}$ and $sk$, and judge if a suspect encoder $F'$ is a copy by: 
\begin{equation*}
    \texttt{Verify}(F') = \left\{
                \begin{array}{ll}
                1, & {{\rm WR} >th_v}\\
                0, & {\text{otherwise}}\\
                \end{array} \right.,
\end{equation*}
here we adopt watermark rate (WR) as the metric to denote the ratio of the verified samples whose outputs $sk'_x$ are close to $sk$.
Concretely, WR is defined as:
\begin{equation*}
    {\rm WR} = \frac{1}{|D_v|}\sum_{x \in D_v}\mathds{1}({\rm sim}(sk'_x,sk)>th_w).
\end{equation*}
In summary, we need two thresholds here: $th_v$ and $th_w$.
$th_w$ is used to calculate WR, and $th_v$ is a threshold to verify the copyright.
We set $th_w = 0.5$ and $th_v = 0.5$ by default.
Note that the $th_w$ can be set to a smaller value as we show in \autoref{section:evaluation} that the WR is 0 for the clean encoders.
The overview of \SSLGuard is depicted in \autoref{figure:overview}.
Concretely, we first train a watermarked encoder that contains the information of the verification dataset and the secret vector.
The clean encoder serves as a query-based API to guide the training process.
The shadow encoder is used to simulate the model stealing process to better preserve the watermark under model stealing attacks.
The watermarked encoder should keep the utility of the clean encoder while preserving the watermark injected in it.

\begin{table}[!t]
\centering
\caption{List of notations.}
\label{table:notation}
\begin{tabular}{r l }
\toprule
{\bf Notation} & {\bf Description} \\
\midrule
$F$, $F_*$, $F_s$ & Clean/Watermarked/Shadow encoder \\
$\mathcal{D}_{t}$, $\mathcal{D}_{s}$ & Target/Shadow dataset \\
$\mathcal{D}_{p}$, $\mathcal{D}_{v}$ & Private/Verification dataset \\
$T$, $M$ & Trigger, Mask \\
$\kappa$, $G$ &Key-tuple, Decoder \\
 $sk$, $sk_x'$ & Secret vector, Decoded vector \\
DA & Downstream accuracy \\
WR & Watermark rate \\
\bottomrule
\end{tabular}
\end{table}

\subsection{Preparation}

To watermark a pre-trained encoder, the defender should prepare a private dataset $\mathcal{D}_{p}$, a mask $M$, and a random trigger $T$.
The mask $M$ is a binary matrix that contains the position information of trigger $T$, which means $M$ and $T$ have the same size as the private samples $x_p$.
Following~\cite{YLZZ19,GWZZLSW21}, we inject the trigger into $x_p$ by:
$$
 \mathcal{P}(x_p, T) =(1-M) \circ x_p + M \circ T, x_p \in \mathcal{D}_{p},
$$
where $\circ$ denotes the element-wise product.
Therefore, given the trigger $T$, we can generate the verification dataset as: $$\mathcal{D}_v=\{x_v | x_v=\mathcal{P}(x_p, T), x_p \in \mathcal{D}_p\}.$$

Here we define three loss functions, i.e., \emph{correlated loss} $\mathcal{L}_{corr}$, \emph{uncorrelated loss} $\mathcal{L}_{uncorr}$, and \emph{embedding match loss} $\mathcal{L}_{match}$ to achieve three goals.
Our first goal is to let the decoded vectors transformed from the verification dataset $\mathcal{D}_v$ to be similar to the secret vector $sk$, and we define \emph{correlated loss function} as:
\begin{equation}
\label{equ:2}
\mathcal{L}_{corr}(\mathcal{D}_v,E) =\frac{-\sum_{x \sim \mathcal{D}_v}{{\rm sim}(sk_x',sk)}}{|\mathcal{D}_v|},
\end{equation}
where ${\rm sim}(\cdot,\cdot)$ is a similarity function.
If not otherwise specified, we use cosine similarity as the similarity function.
The goal of $\mathcal{L}_{corr}$ is to train an encoder and an decoder together to transform $x$ into $sk'_x$, where $sk'_x$ is correlated with $sk$.
The more similar $sk'_x$ and $sk$ are, the smaller $\mathcal{L}_{corr}$ will be.

Secondly, given a clean dataset $\mathcal{D}$ and an encoder $E$, the decoder $G$ transforms embeddings to the orthogonal direction of $sk$ for uncorrelated samples $x \in \mathcal{D}$.
Therefore, we could get another loss function, \emph{uncorrelated loss function}, as:
\begin{equation}
\label{equ:3}
\mathcal{L}_{uncorr}(\mathcal{D}, E) =(\frac{\sum_{x \sim \mathcal{D}}{{\rm sim}(sk'_x,sk)}}{|\mathcal{D}|})^2.
\end{equation}

Finally, we here define an \emph{embedding match loss function} to match the embeddings generated from two encoders $E'$ and $E''$:
\begin{equation}
\label{equ:4}
\mathcal{L}_{match}(\mathcal{D},E',E'') =\frac{-\sum_{x \sim \mathcal{D}}{{\rm sim}(E'(x),E''(x))}}{|\mathcal{D}|}.
\end{equation}
\SSLGuard leverages $\mathcal{L}_{match}$ to maintain the utility of the watermarked encoder and simulate the model stealing attacks.

\subsection{Watermark Injection}

As shown in \autoref{figure:overview}, \SSLGuard adopts three encoders: a clean encoder $F(x;\theta)$, a watermarked encoder $F_*(x;\theta_w)$ and a shadow encoder $F_s(x;\theta_s)$.
Meanwhile, \SSLGuard also uses three datasets: a target dataset $\mathcal{D}_{t}$, a shadow dataset $\mathcal{D}_{s}$, and a verification dataset $\mathcal{D}_{v}$.
In the following part, we will introduce our loss functions for each module.

\mypara{Shadow Encoder}
For the shadow encoder, its task is to mimic the model stealing attacks.
Here we use $\mathcal{D}_{s}$ to simulate the query process.
The loss function of the shadow encoder is:
\begin{equation}
\label{equ:5}
\mathcal{L}_{s} = \mathcal{L}_{match}(\mathcal{D}_{s}, F_*,F_s).
\end{equation}

\mypara{Trigger and Decoder}
Given a verification dataset, we aim to optimize a trigger $T$ and a decoder $G$ to extract $sk$ from both the watermarked encoder and the shadow encoder, but not the clean encoder.
The corresponding loss can be defined as:
\begin{equation}
\label{equ:6}
\mathcal{L}_1 =
\mathcal{L}_{uncorr}(\mathcal{D}_{v},F) + \mathcal{L}_{corr}(\mathcal{D}_v,F_*) + \mathcal{L}_{corr}(\mathcal{D}_v,F_s).
\end{equation}

Besides, for the clean encoder $F$, watermarked encoder $F_*$, and the shadow encoder $F_s$, the decoder should not map the decoded keys closely to $sk$ from the target dataset, the loss to achieve this goal can be defined as: 
\begin{equation}
\label{equ:7}
\mathcal{L}_2 =
 \mathcal{L}_{uncorr}(\mathcal{D}_{t},F) + \mathcal{L}_{uncorr}(\mathcal{D}_{t},F_*) + \mathcal{L}_{uncorr}(\mathcal{D}_{t},F_s).
\end{equation}

Given the above losses, the final loss function for trigger and decoder can be defined as:
\begin{equation}
\label{equ:8}
\mathcal{L}_{TD} = \mathcal{L}_1 + \mathcal{L}_2.
\end{equation}

\mypara{Watermarked Encoder}
For the watermarked encoder, we want it to keep the utility of the clean encoder.
Therefore, for the samples from $\mathcal{D}_{t}$, we force the embeddings from $F$ and $F_*$ to become similar through $\mathcal{L}_{match}$.
The loss $\mathcal{L}_3$ can be defined as:
\begin{equation}
\label{equ:9}
\mathcal{L}_3 = \mathcal{L}_{match}(\mathcal{D}_{t}, F,F_*).
\end{equation}

Meanwhile, the decoder $G$ should successfully extract $sk$ from the verification dataset $\mathcal{D}_{v}$ instead of the target dataset $\mathcal{D}_{t}$.
The corresponding loss $\mathcal{L}_1$ to achieve this goal is defined as:
\begin{equation}
\label{equ:10}
\mathcal{L}_4 = \mathcal{L}_{uncorr}(\mathcal{D}_{t}, F_*) + \mathcal{L}_{corr}(\mathcal{D}_v, F_*).
\end{equation}
The final loss function for the watermarked encoder is:
\begin{equation}
\label{equ:11}
\mathcal{L}_w = \mathcal{L}_3 + \mathcal{L}_4.
\end{equation}

\mypara{Optimization Problem}
After designing all loss functions, we formulate \SSLGuard as an optimization problem.
Concretely, we update the parameters as follows:
\begin{equation}
\label{equ:op}
\begin{split}
    \theta_s & \leftarrow {\rm Optimizer}(\theta_s, \bigtriangledown_{\theta_s}\mathcal{L}_{s}, \eta_s), \\
    T,G & \leftarrow {\rm Optimizer}(T,G, \bigtriangledown_{ T,G}\mathcal{L}_{TD}, \eta_{TD}),\\
    \theta_w & \leftarrow {\rm Optimizer}(\theta_w, \bigtriangledown_{\theta_w}\mathcal{L}_{w}, \eta_w), \\
\end{split}
\end{equation}
where $\eta_s$, $\eta_{TD}$, and $\eta_w$ are learning rates of shadow encoder, watermarked encoder, trigger, and decoder, respectively.
We note that we update $\theta_s$, $T$, $G$, and $\theta_w$ sequentially in one iteration, and we stop the optimization until the iteration reaches the max iteration number.

\section{Evaluation}
\label{section:evaluation}

\subsection{Experimental Setup}
\label{subsection:experimental_setup}

\mypara{Datasets}
We use the following 7 datasets to conduct our experiments.

\begin{itemize}
    \item {\bf ImageNet~\cite{RDSKSMHKKBBF15}.} The ImageNet dataset contains 1.2 million training images distributed in 1,000 classes. Each image has size $224 \times 224 \times 3$.
    \item {\bf CIFAR-10~\cite{CIFAR}} The CIFAR-10 dataset has $60,000$ images in $10$ classes. Among them, there are $50,000$ images for training and $10,000$ images for testing. The size of each image is $32 \times 32 \times 3$.
    \item {\bf CIFAR-100~\cite{CIFAR}.} Similar to CIFAR-10, The CIFAR-100 dataset contains $60,000$ images with size $32 \times 32 \times 3$ in $100$ classes, and there are 500 training images and 100 testing images in each class.
    \item {\bf STL-10~\cite{CNL11}.} The STL-10 dataset consists of $5,000$ training images and $8,000$ testing images in 10 classes. Besides, it also contains $100,000$ unlabeled images. Note that the images on STL-10 are acquired from labeled images on ImageNet.\footnote{\url{https://cs.stanford.edu/~acoates/stl10/}} The size of each image is $96 \times 96 \times 3$.
    \item {\bf GTSRB~\cite{SSSI11}.} German Traffic Sign Recognition Benchmark (GTSRB) contains $39,209$ training images and $12,630$ testing images. It contains $43$-category traffic signs.
    \item {\bf MNIST~\cite{MNIST}.} MNIST is a handwritten digits dataset that contains 60,000 training images and 10,000 testing images in 10 classes. Each image has size $28 \times 28 \times 1$.
    \item {\bf FashionMNIST~\cite{XRV17}.} FashionMNIST (F-MNIST) is a Zalando's article image dataset that has 10 classes. It has 60,000 training images and 10,000 testing images. Each sample is a grayscale image with size $28 \times 28 \times 1$.
\end{itemize}

We resize images of all datasets to $224 \times 224 \times 3$ in our experiments.
We use ImageNet as the pre-training dataset; STL-10, CIFAR-10, F-MNIST, and MNIST as the downstream datasets; and STL-10, CIFAR-10, CIFAR-100, and GTSRB as the query dataset (to launch model stealing attacks).
Note that for the STL-10 dataset, we randomly split the unlabeled samples (100,000) of it into two parts (each containing 50,000 samples).
We consider the first part as the unlabeled STL-10 dataset and the second part as the same distribution unlabeled STL-10 dataset which is denoted as STL-10 (s).

\mypara{Pre-trained Encoder}
In our experiments, we adopt real-world contrastive learning pre-trained encoders as the victim encoders.
Concretely, we download the checkpoints of the encoders from the official website (i.e., SimCLR\footnote{\url{https://github.com/google-research/simclr}} and MoCo v2\footnote{\url{https://github.com/facebookresearch/moco}}) or the public platform (i.e., BYOL\footnote{\url{https://github.com/yaox12/BYOL-PyTorch}}).
All the encoders are ResNet-50 pre-trained on ImageNet.

\mypara{Downstream Classifier}
We use a $3$-layer MLP as the downstream classifier with $512$ and $256$ neurons in its hidden layer.
For each downstream task, we freeze the parameters of the pre-trained encoders and train the downstream classifier for $20$ epochs using Adam optimizer~\cite{KB15} with $0.005$ learning rate.

\mypara{SSLGuard}
We reload the clean encoder and fine-tune it to be the watermarked encoder.
Note that we freeze the weights in batch normalization layers following the settings by Jia et al.~\cite{JLG22}.
We consider the unlabeled STL-10 dataset (with only 50,000 images as mentioned above)  as both $\mathcal{D}_{s}$ and $\mathcal{D}_{t}$, and adopt a ResNet-50 as the shadow encoder's architecture.
We sample 100 images from 5 random classes on ImageNet as our $\mathcal{D}_{p}$.
Note that each class contains 20 images and the $\mathcal{D}_{p}$ for watermarking SimCLR, MoCo v2, and BYOL are non-overlapping.
For each sample in $\mathcal{D}_{p}$, $35\%$ space will be patterned by the trigger.
We leverage the SGD optimizer with $0.01$ learning rate to train both the watermarked encoder and shadow encoder for 50 epochs.
The batch size in our experiment is $8$.
The dimension of $sk$ is $256$.
For the trigger, we randomly generate a $224 \times 224 \times 3$ tensor from a uniform distribution in $[0,1]$ as the initial trigger.
We use a 3-layer MLP as the decoder $G$.
The numbers of $G$'s neurons are 512, 256, and 256, respectively.
We use the SGD optimizer with $0.005$ learning rate to update both the decoder and the trigger.

\subsection{Clean Downstream Accuracy}
\label{subsection:clean_pretrained_encoder}

Given three \emph{clean} SSL pre-trained encoders (i.e., pre-trained by SimCLR, MoCo v2, and BYOL on ImageNet), we first measure their downstream accuracy, denoted as \emph{clean downstream accuracy (CDA)}, for different tasks.
We consider $4$ downstream classification tasks, i.e., STL-10, CIFAR-10, MNIST, and F-MNIST.
The CDA are shown in \autoref{table:clean_DA}.
We observe that the SSL pre-trained encoders can achieve remarkable performance on different downstream tasks, which means the SSL pre-trained encoders can learn high-level semantic information from one task (i.e., ImageNet), and the informative embeddings can generalize to other tasks (i.e., STL-10 and CIFAR-10).
Meanwhile, the cost of pre-training SSL encoders is expensive (see \autoref{table:money}), such observation further demonstrates the necessity of protecting the copyright of the SSL pre-trained encoders.
Note that we adopt CDA as our baseline accuracy.
Later we measure an encoder's performance by comparing its DA with CDA.

\begin{table}[ht]
\centering
\caption{Clean downstream accuracy (CDA).}
\label{table:clean_DA}
\begin{tabular}{l| ccc }
\toprule
Downstream Task & SimCLR & MoCo v2 & BYOL \\
\midrule
STL-10    & 0.783  & 0.889 & 0.948 \\
CIFAR-10  & 0.766  & 0.712 & 0.855 \\
MNIST     & 0.974  & 0.940 & 0.974 \\
F-MNIST   & 0.874  & 0.852 & 0.894 \\
\bottomrule
\end{tabular}
\end{table}

\subsection{Model Stealing Attacks}

Since the SSL pre-trained encoders (clean encoders) are powerful, we then evaluate whether they are vulnerable to model stealing attacks.
To build a surrogate encoder, we consider three key information, i.e., the surrogate encoder's architecture, the distribution of the query dataset, and the similarity function used to ``copy'' the victim encoder.

\begin{figure*}[!t]
\centering
\begin{subfigure}{0.65\columnwidth}
\includegraphics[width=\columnwidth]{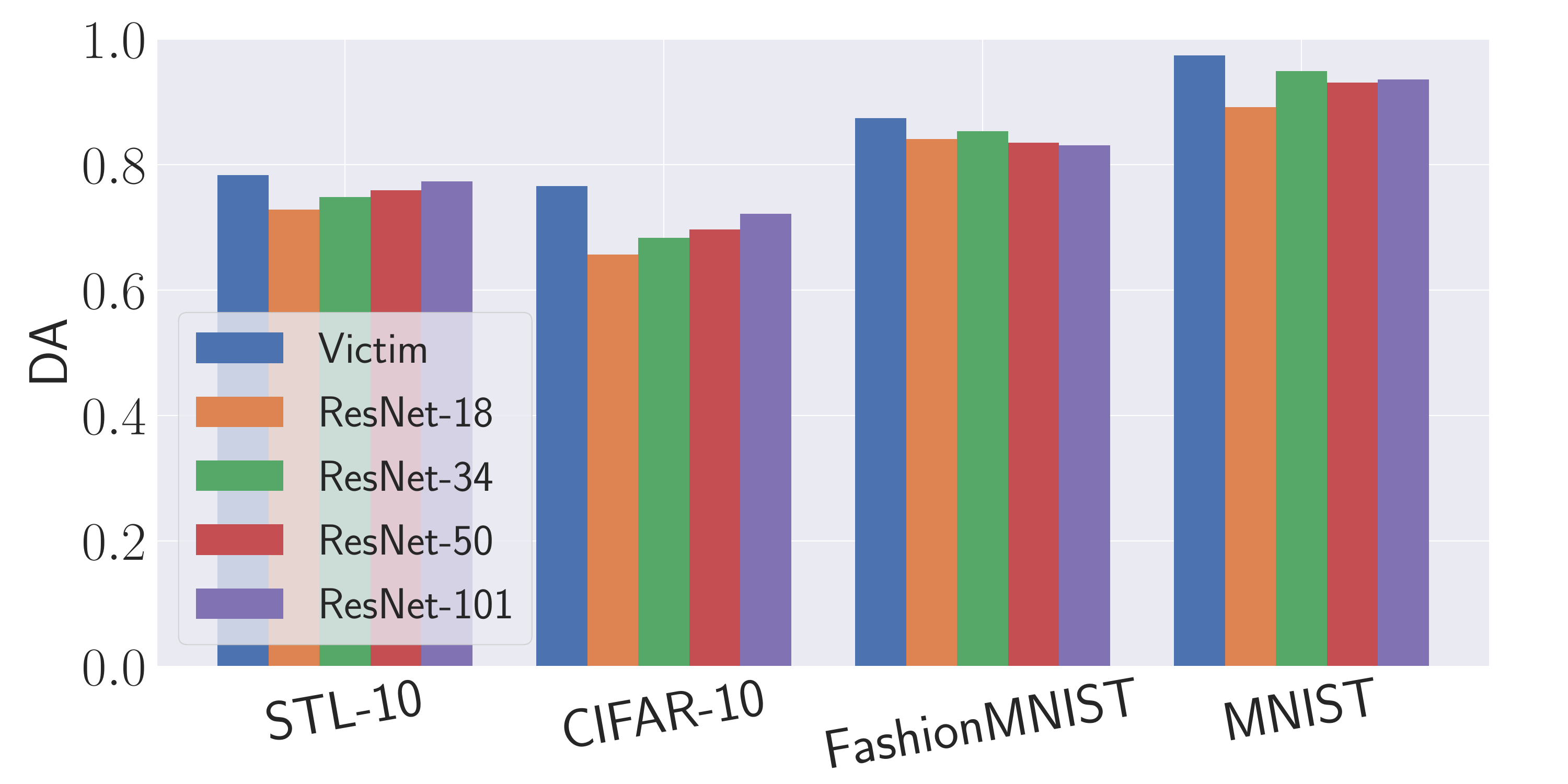}
\caption{SimCLR}
\label{fig:steal_clean_arch_a}
\end{subfigure}
\begin{subfigure}{0.65\columnwidth}
\includegraphics[width=\columnwidth]{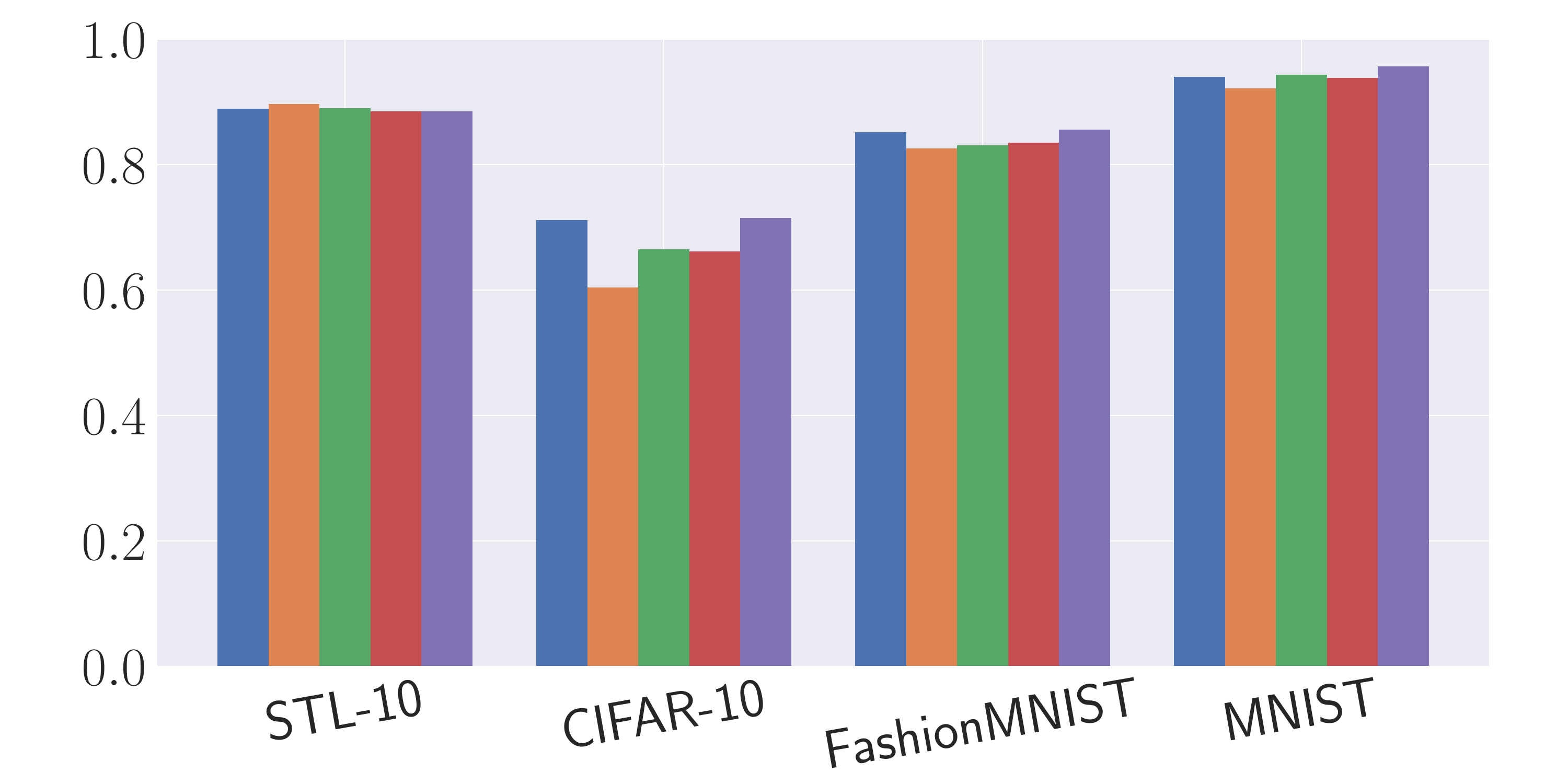}
\caption{MoCo v2}
\label{fig:steal_clean_arch_b}
\end{subfigure}
\begin{subfigure}{0.65\columnwidth}
\includegraphics[width=\columnwidth]{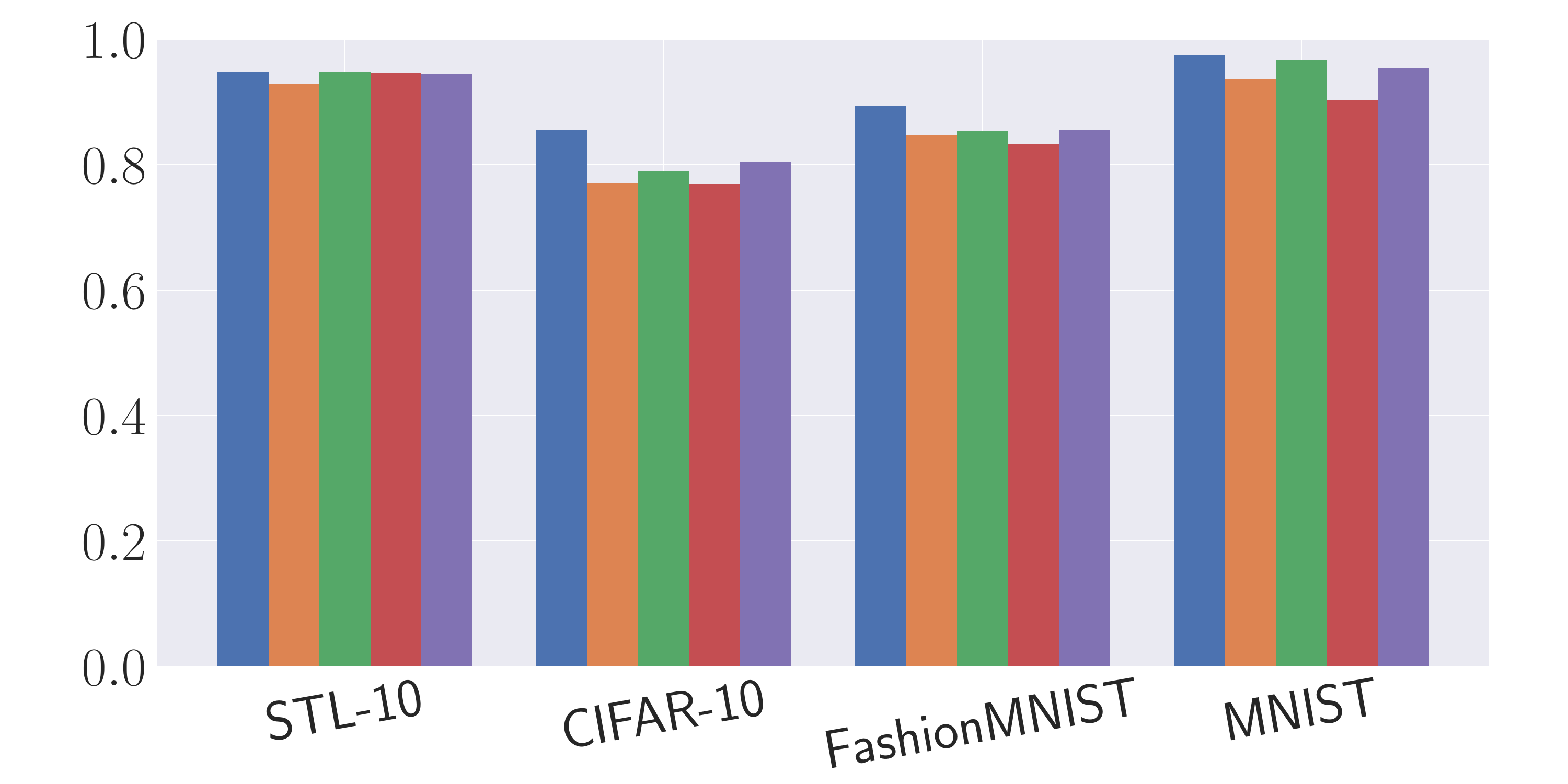}
\caption{BYOL}
\label{fig:steal_clean_arch_c}
\end{subfigure}
\caption{The performance of surrogate encoders trained with different architectures.}
\label{fig:steal_clean_arch}
\end{figure*}

\begin{figure*}[!t]
\centering
\begin{subfigure}{0.65\columnwidth}
\includegraphics[width=\columnwidth]{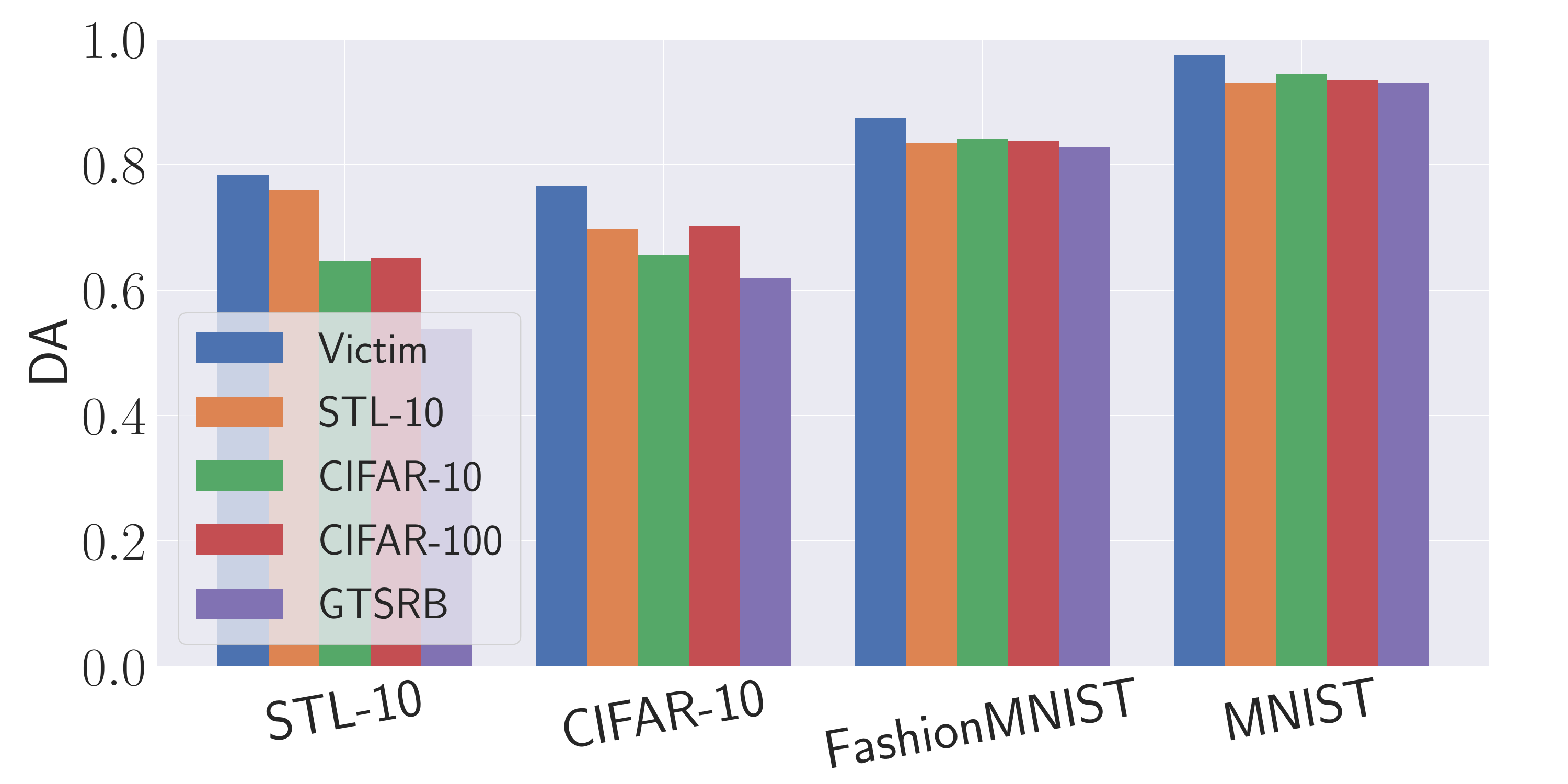}
\caption{SimCLR}
\label{fig:steal_clean_data_a}
\end{subfigure}
\begin{subfigure}{0.65\columnwidth}
\includegraphics[width=\columnwidth]{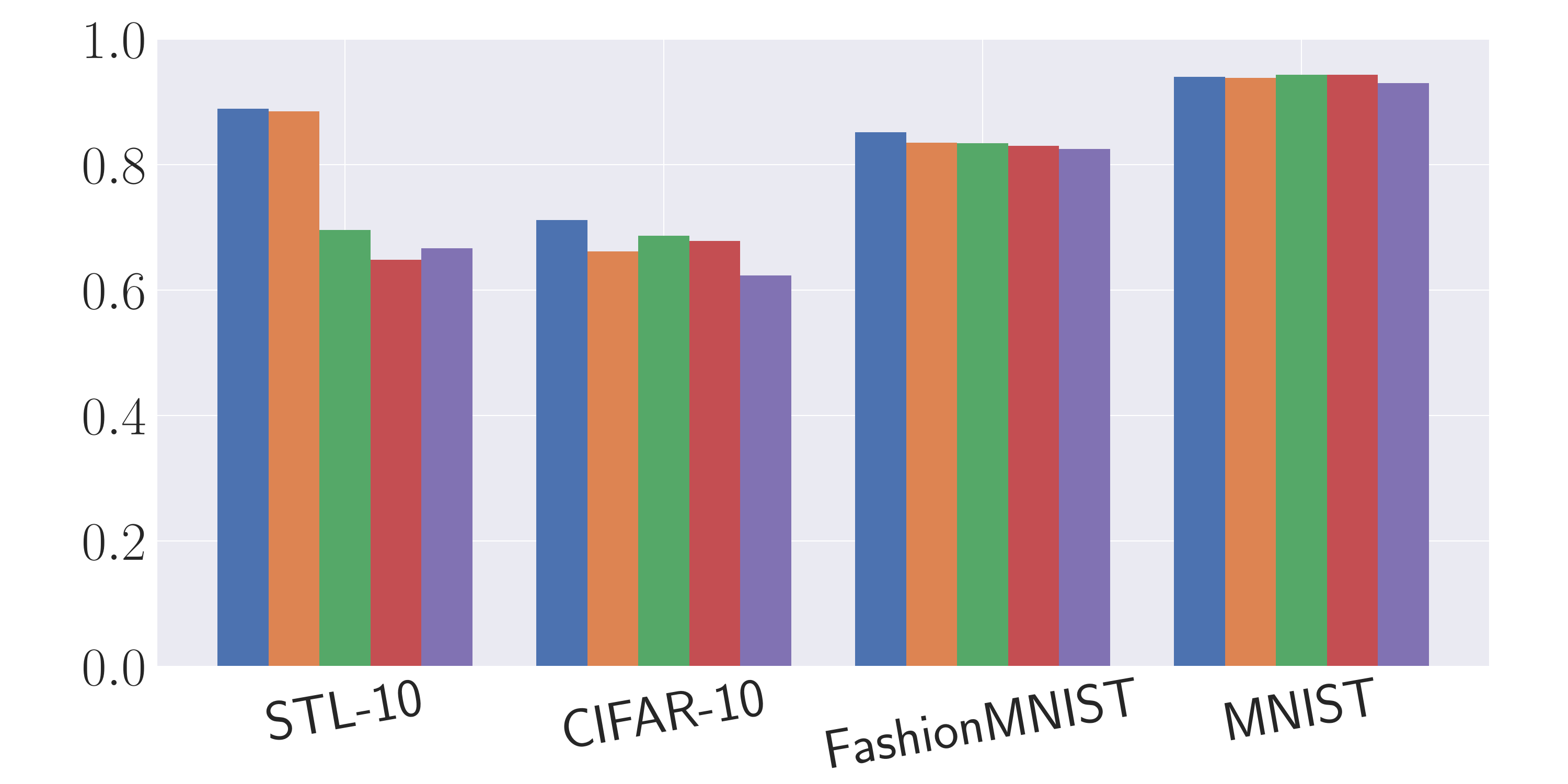}
\caption{MoCo v2}
\label{fig:steal_clean_data_b}
\end{subfigure}
\begin{subfigure}{0.65\columnwidth}
\includegraphics[width=\columnwidth]{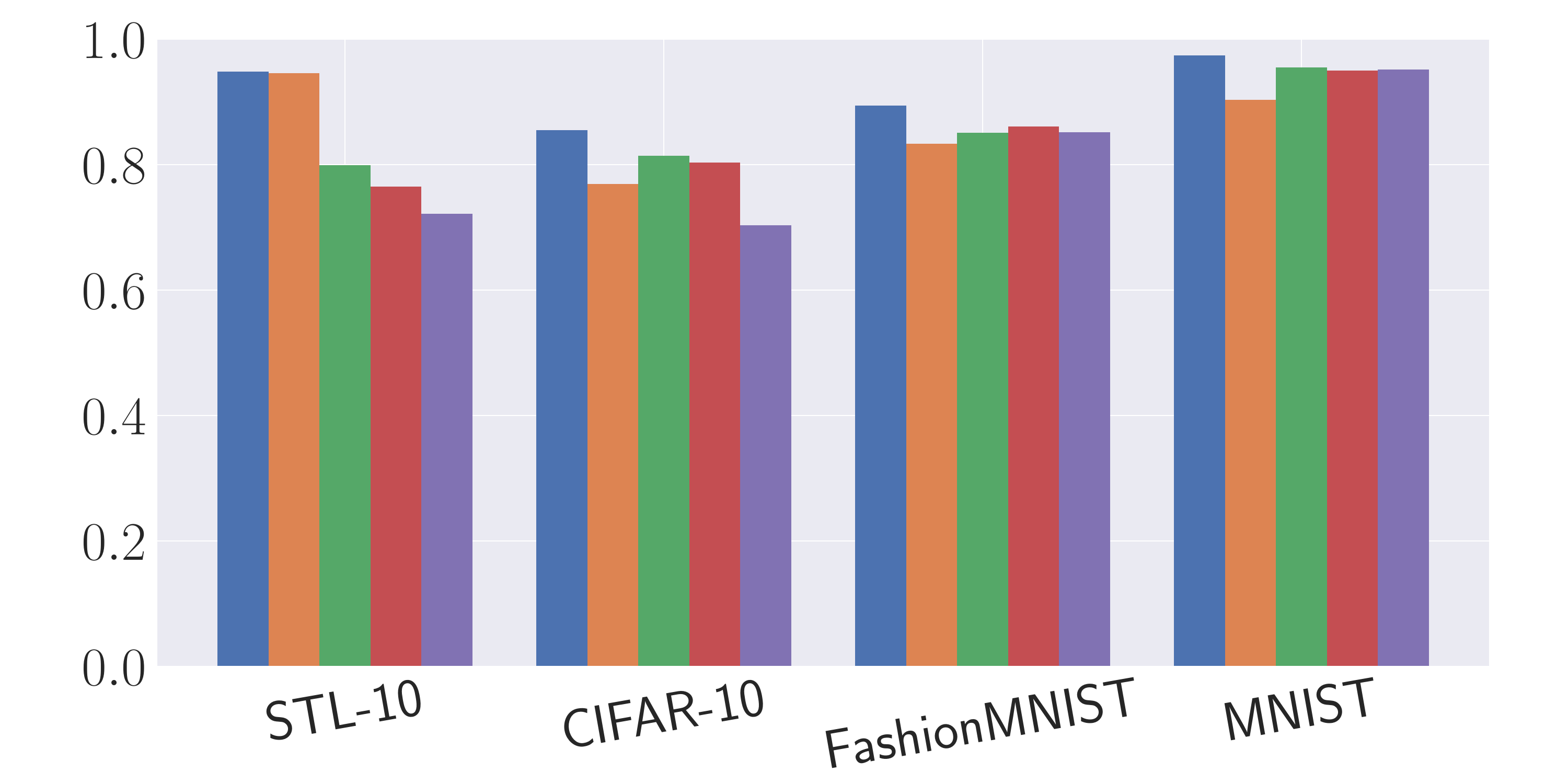}
\caption{BYOL}
\label{fig:steal_clean_data_c}
\end{subfigure}
\caption{The performance of surrogate encoders trained with different query datasets.}
\label{fig:steal_clean_data}
\end{figure*}

\begin{figure*}[!t]
\centering
\begin{subfigure}{0.65\columnwidth}
\includegraphics[width=\columnwidth]{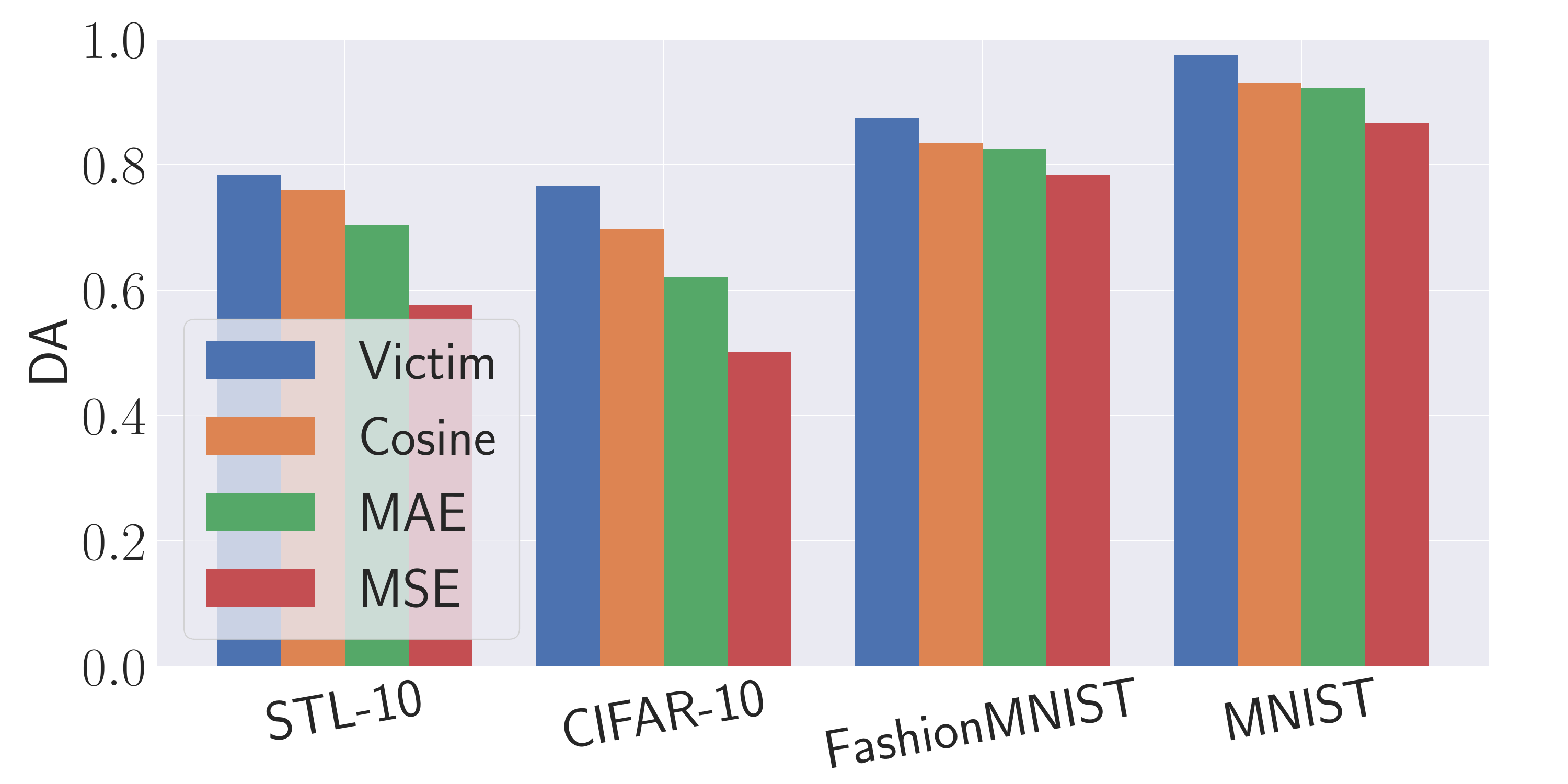}
\caption{SimCLR}
\label{fig:steal_clean_sim_a}
\end{subfigure}
\begin{subfigure}{0.65\columnwidth}
\includegraphics[width=\columnwidth]{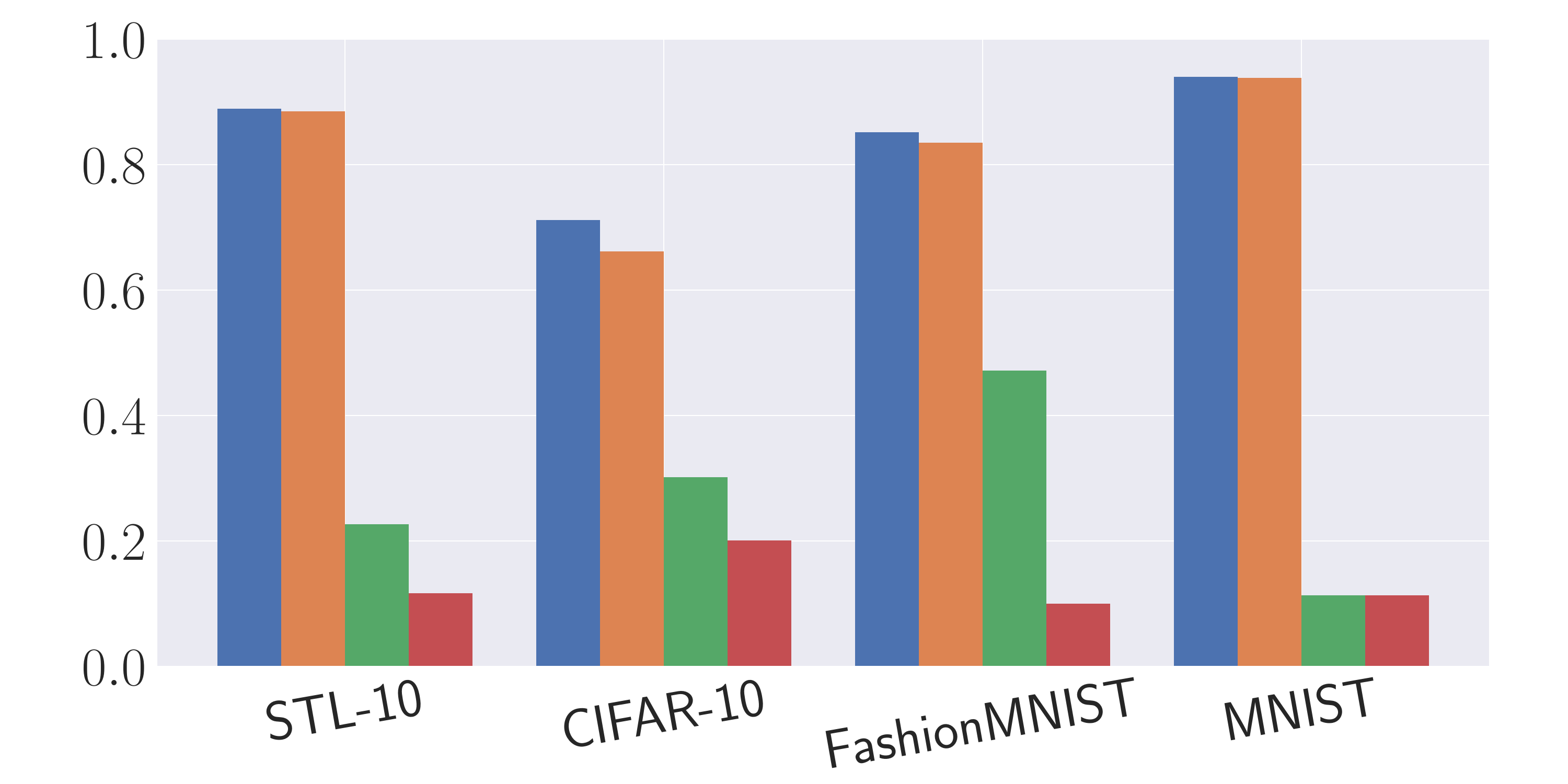}
\caption{MoCo v2}
\label{fig:steal_clean_sim_b}
\end{subfigure}
\begin{subfigure}{0.65\columnwidth}
\includegraphics[width=\columnwidth]{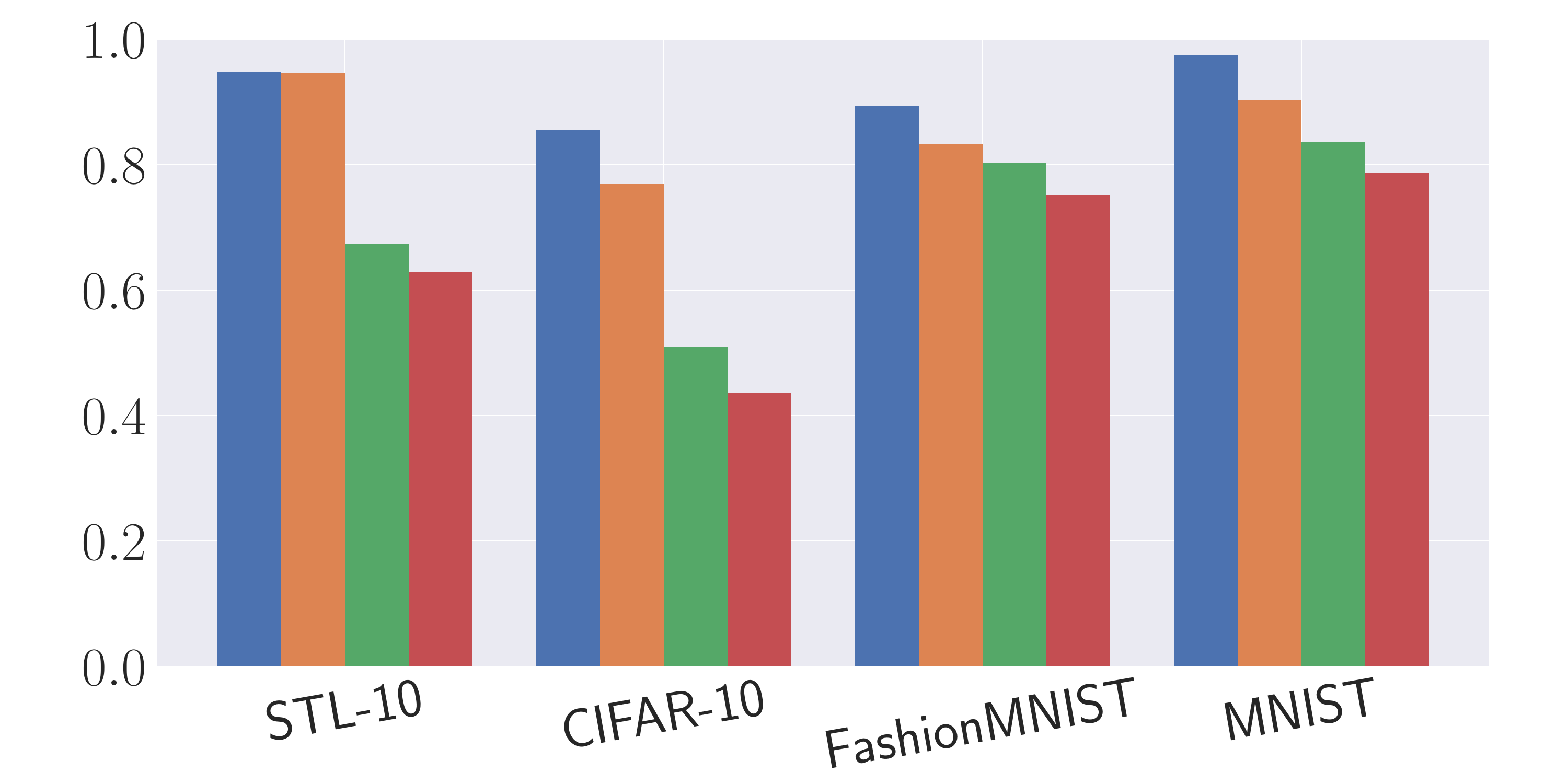}
\caption{BYOL}
\label{fig:steal_clean_sim_c}
\end{subfigure}
\caption{The performance of surrogate encoders trained with different loss functions.}
\label{fig:steal_clean_sim}
\end{figure*}

\mypara{Surrogate Encoder's Architecture}
We first investigate the impact of the surrogate encoder's architecture.
Note that here we adopt the unlabeled STL-10 dataset (with 50,000 unlabeled samples) as the query dataset and cosine similarity as the similarity function to measure the difference between the victim and surrogate encoders' embeddings.
Since the architecture of the victim encoder can be non-public, the adversaries may try different surrogate encoder architectures to perform the model stealing attacks.
Concretely, we assume the adversaries may leverage ResNet-18, ResNet-34, ResNet-50, or ResNet-101 as the surrogate encoder's architecture.
If the output dimension is different from ResNet-50 (the architecture of the victim encoder), e.g., ResNet-18/ResNet-34 outputs 512-dimensional embeddings, we leverage an extra linear layer to transform them into 2048-dimension.
The DA of surrogate encoders is summarized in \autoref{fig:steal_clean_arch}.
A general trend is that the deeper the surrogate encoder's architecture, the better performance it can achieve on the downstream tasks.
For instance, for SimCLR (\autoref{fig:steal_clean_arch_a}), the DA on STL-10 and CIFAR-10 are 0.728 and 0.657 when the surrogate encoder's architecture is ResNet-18, while the DA increases to 0.759 and 0.697 when the surrogate encoder's architecture is changed to ResNet-50.
This may be because a deeper model architecture can provide a wider parameter space and greater representation ability.
Therefore, in general, deeper surrogate encoder's architectures can better ``copy'' the functionality from victim encoders.
Note that in the following experiments, the adversary uses ResNet-50 as the surrogate encoder's architecture by default as it has comparable performance to ResNet-101 while requiring fewer resources.

\mypara{Distribution of the Query Dataset}
Secondly, we evaluate the impact of the query dataset's distribution.
In the real-world scenario, the adversary may or may not have the query dataset that is from the same distribution as the victim encoder's pre-training dataset.
Here the adversary leverages ResNet-50 as the surrogate model's architecture and cosine similarity as the similarity function.
Regarding the query dataset, the adversary may leverage the training dataset of CIFAR-10, CIFAR-100, and GTSRB and the unlabeled dataset of STL-10 to perform the attacks.
The results are shown in \autoref{fig:steal_clean_data}.
First, we observe that the model stealing attack is more effective with querying by the same distribution dataset as the pre-training dataset.
For instance, given the victim model trained by SimCLR (\autoref{fig:steal_clean_data_a}), when the downstream task is STL-10 classification, the DA for the surrogate encoders are 0.759, 0.646, 0.651, and 0.538 when the query dataset is STL-10, CIFAR-10, CIFAR-100, and GTSRB, respectively.
This demonstrates that the same distribution query dataset can better steal the functionality of the victim encoder.

Another observation is that the distribution of the surrogate dataset may also influence DA on different tasks.
For instance, given the victim model trained by BYOL (\autoref{fig:steal_clean_data_c}),
when the downstream task is CIFAR-10 classification, the DA is 0.814 with CIFAR-10 as the query dataset, while only 0.769 with STL-10 as the query dataset.
However, when the downstream task is STL-10 classification, the DA is 0.799 with CIFAR-10 as the query dataset but increases to 0.946 with STL-10 as the query dataset.
Therefore, if the adversary is aware of the downstream task, they can construct a query dataset that is close to the downstream task to improve the stealing performance.

\mypara{Similarity Function}
Finally, we investigate the effect of similarity functions used in model stealing attacks.
Besides cosine similarity, the adversary can also use mean absolute error (MAE) and mean square error (MSE) to match the victim encoder's embeddings.
Here we assume that the adversary leverages ResNet-50 as the surrogate model's architecture and STL-10 as the query dataset.
The results are shown in \autoref{fig:steal_clean_sim}.
We can see that cosine similarity outperforms MAE and MSE in most settings.
For instance, given the victim model trained by MoCo v2 (\autoref{fig:steal_clean_sim_b}), the DA are all below 0.5 when using MAE and MSE.
This can be credited to the normalization effect of cosine similarity, which helps to better learn the embeddings~\cite{GSATRBDPGAPKMV20}.
This indicates that cosine similarity may better facilitate the stealing process.

\mypara{Monetary Cost} 
We compare the monetary costs of pre-training an SSL encoder from scratch and stealing an SSL encoder.
We first measure the training cost of the encoders.
To pre-train a ResNet-50 encoder, SimCLR needs 60 hours with 32 TPU v3s, MoCo v2 uses 212 hours with 8 NVIDIA V100 GPUs, and BYOL takes 72 hours with 32 NVIDIA V100 GPUs (the training information is from the official or open-source implementation as mentioned in \autoref{subsection:experimental_setup}).
The cost of model stealing contains two parts: querying the victim encoders and training the surrogate encoders locally.
We use the GPU price from Google cloud\footnote{\url{https://cloud.google.com/compute/gpus-pricing}} to calculate the price for pre-training (i.e., We run our experiments on one NVIDIA A100 GPU whose price is \$2.934 per hour).
Meanwhile, we refer to the querying price, i.e., \$1 per 1,000 queries, from AWS.\footnote{\url{https://aws.amazon.com/rekognition/pricing}}
We adopt the unlabeled STL-10 dataset (50,000 samples), cosine similarity, and different architectures to launch model stealing attacks.
The monetary costs are shown in \autoref{table:money}.
We observe that the cost of stealing the pre-trained encoder is much smaller than pre-training it from scratch.
For instance, pre-trains a BYOL ResNet-50 encoder takes $\$ 5,713.92$ while stealing it with a ResNet-101 encoder only takes $\$ 72.49$.
This indicates that an adversary can ``copy'' the victim encoder with much less cost.

\begin{table}[h]
\centering
\caption{Monetary Cost (\$). Here Res denotes ResNet.}
\label{table:money}
\setlength{\tabcolsep}{1.1mm}{
\begin{tabular}{c|c|cccc}
\toprule
& \multirow{2}{*}{Pre-training} & \multicolumn{4}{c}{Stealing}  \\
&              & Res-18 & Res-34 & Res-50 & Res-101 \\
\midrule
SimCLR   & {\bf 1,920.00} & 58.24 & 61.10 & 66.67 & 74.50 \\
MoCo v2  & {\bf 4,206.08} & 58.13 & 61.09 & 66.55 & 74.37 \\
BYOL     & {\bf 5,713.92} & 58.16 & 60.84 & 64.28 & 72.49 \\
\bottomrule
\end{tabular}
}
\end{table}

\subsection{\SSLGuard}

In this section, we adopt \SSLGuard to inject the watermarks into the clean encoders pre-trained by SimCLR, MoCo v2, and BYOL.
We aim to validate four properties of \SSLGuard, i.e., effectiveness, utility, undetectability, and efficiency.
We will discuss the robustness of \SSLGuard separately in \autoref{sec:robust}.

\mypara{Effectiveness}
We first evaluate the effectiveness of \SSLGuard.
Concretely, we check whether the model owner can extract the watermark from the watermarked encoders.
Ideally, the watermark should be successfully extracted from the watermarked encoder $F_*$ and shadow encoder $F_s$, but not the clean encoder $F$.
We use the generated key-tuple $\kappa$ to measure the watermark rate (WR) for $F$, $F_*$, and $F_s$ on three SSL algorithms.
As shown in \autoref{table:effectiveness}, the WR of $F_*$ and $F_s$ are all 1.00, which means encoder $F_*$ and $F_s$ both contain the information of $\mathcal{D}_v$ and $sk$.
Meanwhile, the WR of $F$ is 0.00.
This means \SSLGuard is generic and does not judge a clean encoder to be a watermarked encoder.

\begin{table}[ht]
\centering
\caption{Effectiveness.}
\label{table:effectiveness}
\begin{tabular}{c|ccc}
\toprule
Encoder & SimCLR & MoCo v2 & BYOL \\
\midrule
$F$       & 0.00  & 0.00 & 0.00  \\
$F_*$     & {\bf 1.00}  & {\bf 1.00} & {\bf 1.00}  \\
$F_s$     & {\bf 1.00}  & {\bf 1.00} & {\bf 1.00}  \\
\bottomrule
\end{tabular}
\end{table}

\begin{figure}[!t]
\centering
\begin{subfigure}{0.4\columnwidth}
\includegraphics[width=\columnwidth]{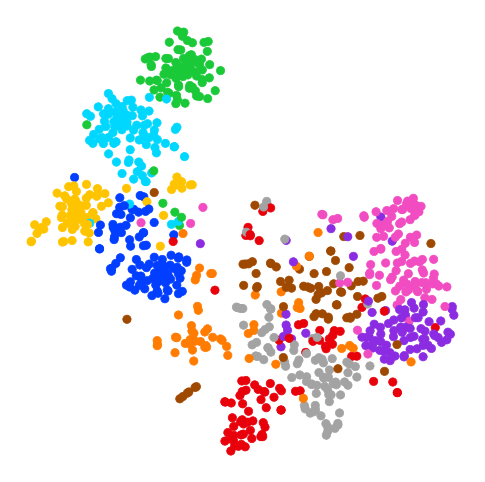}
\caption{$F^{byol}$}
\label{fig:tsne_embed_byol_a}
\end{subfigure}
\begin{subfigure}{0.4\columnwidth}
\includegraphics[width=\columnwidth]{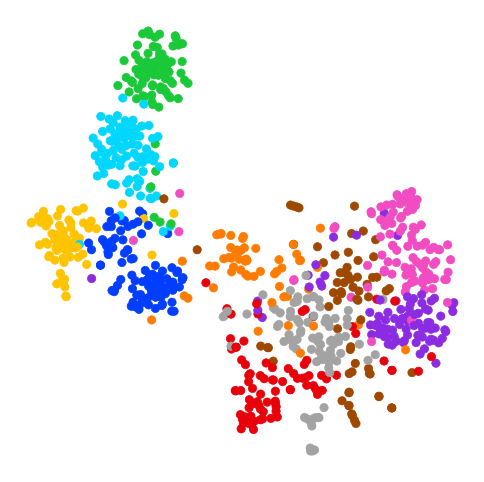}
\caption{$F_*^{byol}$}
\label{fig:tsne_embed_byol_b}
\end{subfigure}
\caption{The t-SNE visualizations of features output from $F^{byol}$ and $F_*^{byol}$ when we input 800 samples in 10 classes randomly chosen from the STL-10 training dataset. Each point represents an embedding. Each color represents one class.}
\label{fig:tsne_embed_byol}
\end{figure}

\mypara{Fidelity}
One of the initial intentions of \SSLGuard is to maintain the utility of the original downstream task.
To verify its fidelity, we first take BYOL as an example and visualize embeddings output from $F^{byol}$ (the clean encoder pre-trained by BYOL) and $F_*^{byol}$ using t-Distributed Neighbor Embedding (t-SNE)~\cite{MH08}, which is depicted in \autoref{fig:tsne_embed_byol}.
We observe that the t-SNE results of $F^{byol}$ and $F_*^{byol}$ are almost identical and the embeddings are successfully separated by both encoders.
This demonstrates that watermarked encoder trained by \SSLGuard can faithfully reproduce the embeddings generated from the clean encoder.
Also, we train downstream classifiers by using three watermarked encoders $F_*^{simclr}$, $F_*^{moco}$ and $F_*^{byol}$ on STL-10, CIFAR-10, F-MNIST, and MNIST.
\autoref{table:utility} shows the DA in different scenarios.
We observe that the DA of the watermarked encoders are almost the same as that of the clean encoders.
For instance, compared to $F^{simclr}$, the DA for $F_*^{simclr}$ only drops up to 0.009 from CDA.
The evaluation shows that \SSLGuard does not sacrifice the utility of the clean encoders.

\begin{table}[ht]
\centering
\caption{Fidelity (DA). The value in the parenthesis denotes the difference between CDA.}
\label{table:utility}
\setlength{\tabcolsep}{1.1mm}{
\begin{tabular}{l|ccc}
\toprule
Task       & $F_*^{simclr}$ & $F_*^{moco}$ & $F_*^{byol}$ \\
\midrule
STL-10     & 0.781 ({\bf -0.002})  & 0.888 ({\bf -0.001}) & 0.940 ({\bf -0.008}) \\
CIFAR-10   & 0.765 ({\bf -0.001})  & 0.701 ({\bf -0.011}) & 0.857 ({\bf +0.002})  \\
MNIST      & 0.965 ({\bf -0.009})  & 0.956 ({\bf +0.016}) & 0.966 ({\bf +0.002}) \\
F-MNIST & 0.878 ({\bf +0.004})  & 0.845 ({\bf -0.007}) & 0.894 ({\bf +0.000}) \\
\bottomrule
\end{tabular}
}
\end{table}

\mypara{Undetectability}
We then check if the watermark can be extracted by a \emph{no-matching} key-tuple.
Through \SSLGuard, we generate three key-tuples: $\kappa^{simclr}$, $\kappa^{moco}$  and $\kappa^{byol}$.
We use one of the key-tuples to verify other watermarked encoders, such as using $\kappa^{simclr}$ to judge $F_*^{moco}$.
As shown in \autoref{table:undetectability}, we see that the WR are all 0.00 in \emph{no-match} pairs, which means we cannot use a non-matching $\kappa$ to verify a watermarked encoder.

\begin{table}[ht]
\centering
\caption{Undetectability.}
\label{table:undetectability}
\begin{tabular}{c|ccc}
\toprule
Key-tuple         & $F_*^{simclr}$ & $F_*^{moco}$ & $F_*^{byol}$ \\
\midrule
$\kappa^{simclr}$ & {\bf 1.00} & 0.00 & 0.00 \\
$\kappa^{moco}$   & 0.00 & {\bf 1.00} & 0.00 \\
$\kappa^{byol}$   & 0.00 & 0.00 & {\bf 1.00} \\
\bottomrule
\end{tabular}
\end{table}

\mypara{Efficiency}
\SSLGuard injects watermark into SimCLR, MoCo v2, and BYOL using 17.5hrs, 17.36hrs, and 10.70hrs, respectively, which are only 29.17\%, 8.19\%, and 14.86\% of the time cost to pre-train SSL encoders, and the watermark extraction time is only 1.51s, 2.08s, and 1.82s, respectively.
Note also that we use only a single GPU (A100) in the watermark injection process, which is much less than the requirement for pre-training the SSL encoders.
This demonstrates that \SSLGuard can inject and extract watermarks efficiently.

\begin{figure*}[t]
\centering
\begin{subfigure}{0.65\columnwidth}
\includegraphics[width=\columnwidth]{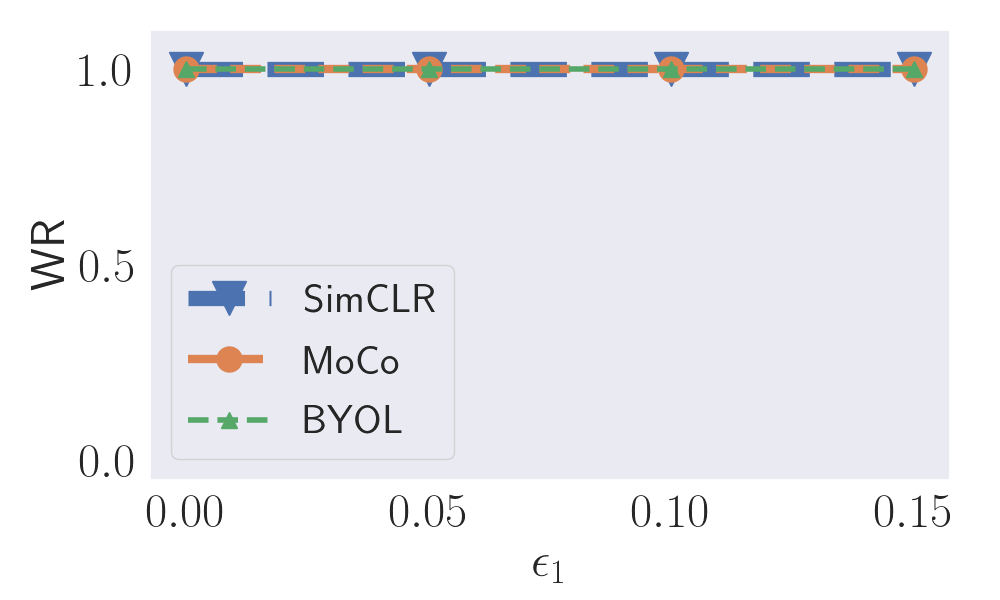}
\caption{Input noising}
\label{fig:robust_wr_a}
\end{subfigure}
\begin{subfigure}{0.65\columnwidth}
\includegraphics[width=\columnwidth]{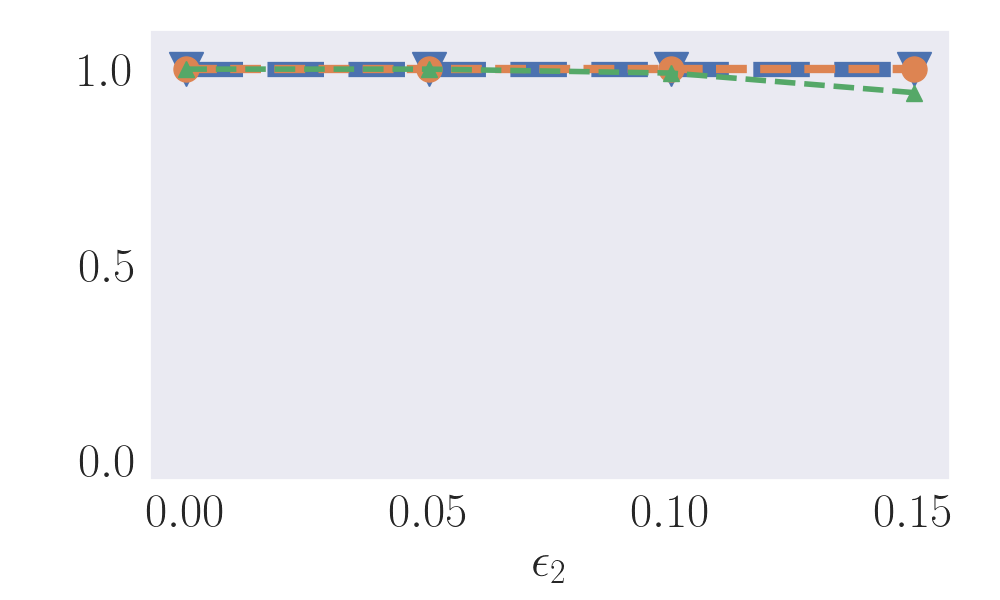}
\caption{Output noising}
\label{fig:robust_wr_b}
\end{subfigure}
\begin{subfigure}{0.65\columnwidth}
\includegraphics[width=\columnwidth]{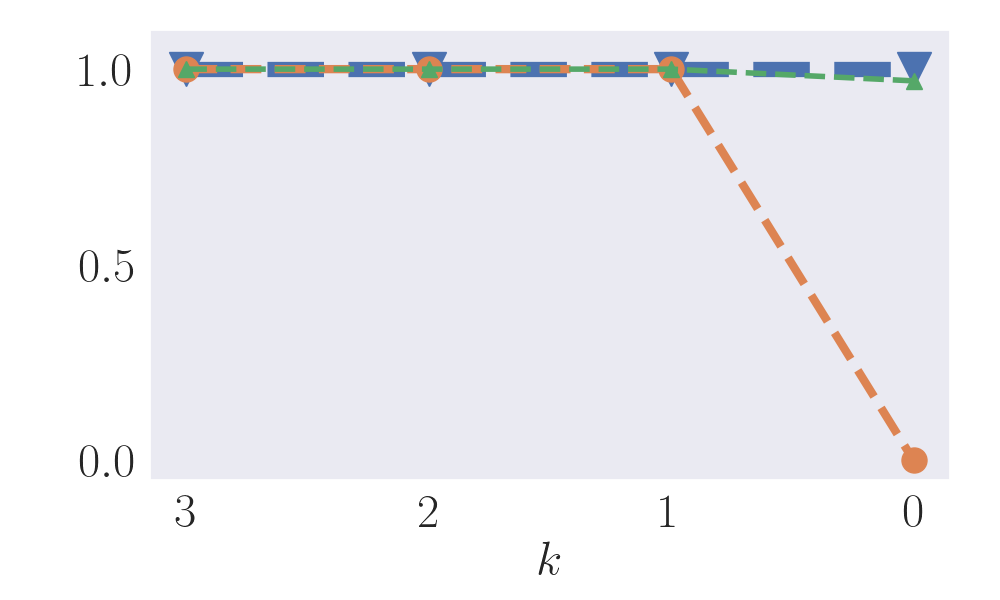}
\caption{Output truncation}
\label{fig:robust_wr_c}
\end{subfigure}
\caption{The WR on different watermark removal attacks.}
\label{fig:robust_wr}
\end{figure*}

\begin{figure*}[t]
\centering
\begin{subfigure}{0.65\columnwidth}
\includegraphics[width=\columnwidth]{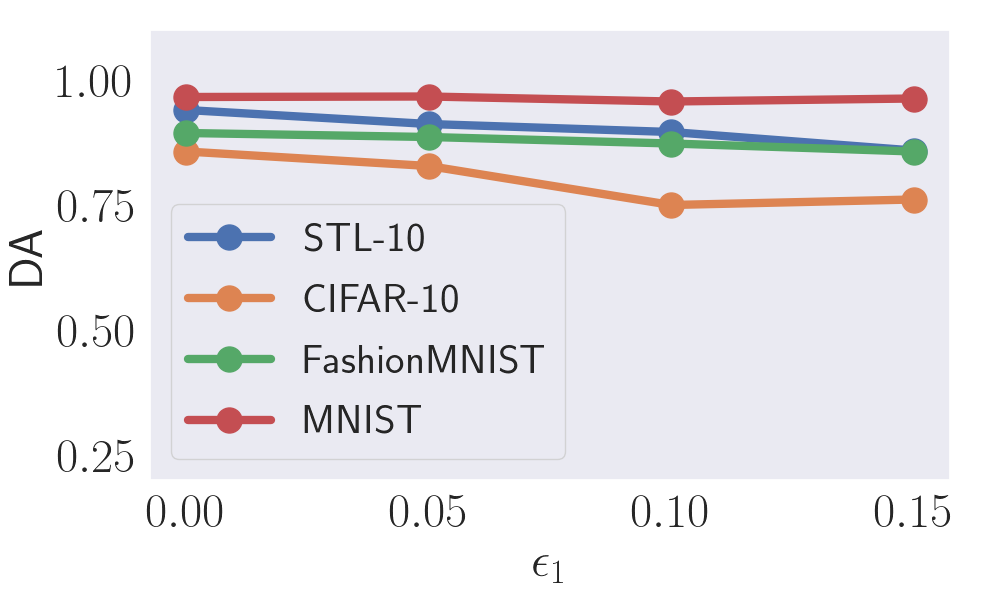}
\caption{Input noising}
\label{fig:robust_da_a}
\end{subfigure}
\begin{subfigure}{0.65\columnwidth}
\includegraphics[width=\columnwidth]{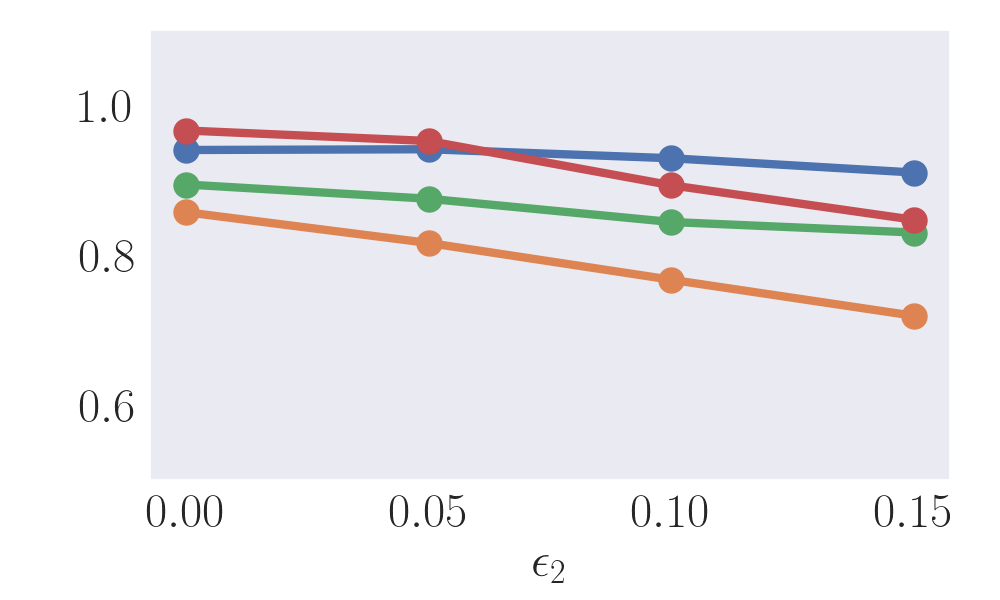}
\caption{Output noising}
\label{fig:robust_da_b}
\end{subfigure}
\begin{subfigure}{0.65\columnwidth}
\includegraphics[width=\columnwidth]{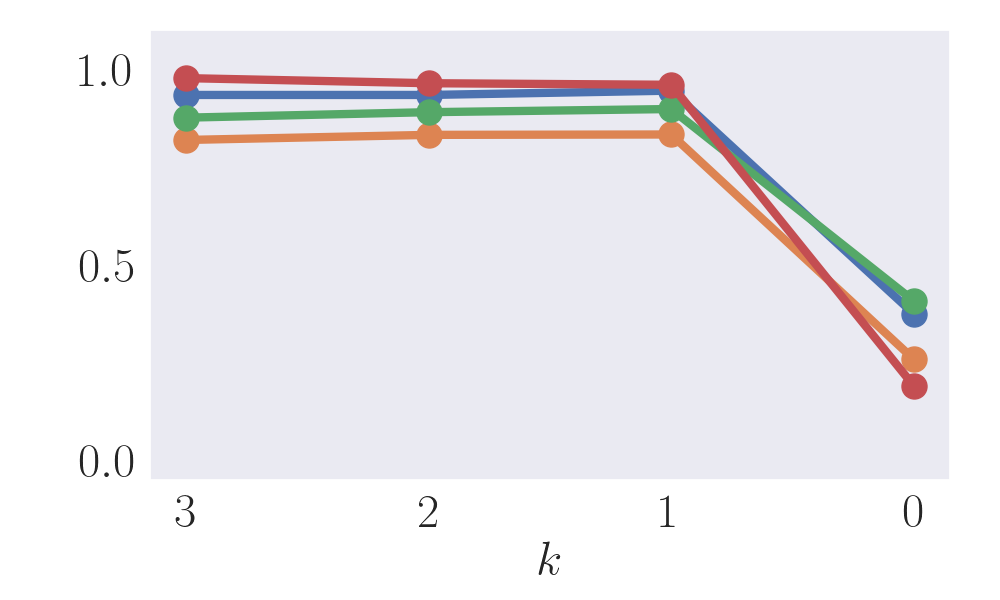}
\caption{Output truncation}
\label{fig:robust_da_c}
\end{subfigure}
\caption{The DA on different watermark removal attacks. The victim encoder is BYOL.}
\label{fig:robust_da}
\end{figure*}

\subsection{Robustness}
\label{sec:robust}

We now quantify the robustness of \SSLGuard.
Concretely, we evaluate \SSLGuard against model stealing and the following watermark removal attacks: Input preprocessing, output perturbing, and model modification.
For instance, the adversary can add noise to the input samples or output embeddings.
Also, the adversary can modify the parameters of the encoder by overwriting, pruning, and fine-tuning.
Since watermark removal attacks may affect the performance of the encoders, and the adversary aims to "clean" the encoder but keep its functionality, we measure DA and WR simultaneously of these surrogate encoders.
We note that the victim encoders are the watermarked encoders, and we leverage SimCLR, MoCo, and BYOL to denote $F_*^{simclr}$, $F_*^{moco}$, and $F_*^{byol}$ in this subsection.
Regarding the downstream accuracy, we only show the results on BYOL (SimCLR and MoCo have similar trends).

\subsubsection{Input Preprocessing.}

Here we consider that the adversary may add i.i.d.Gaussian noise to each input image by $x'=x+\epsilon_1 \cdot \mathcal{N}(0,1)$.
We evaluate DA on four downstream tasks and WR when we use different $\epsilon_1$.
The results of WR are shown in \autoref{fig:robust_wr_a} and DA are shown in \autoref{fig:robust_da_a}.
We first observe that DA drops as $\epsilon_1$ increases.
For instance, the DA on CIFAR-10 drops from 0.932 to 0.865 when $\epsilon_1$ increases from 0.05 to 0.15.
On the other hand, the WR are all 1.00 for different $\epsilon_1$ on SimCLR, MoCo, and BYOL, respectively.
This may be because when we inject the trigger into $\mathcal{D}_p$, the distribution of $\mathcal{D}_v$ is too special, so our watermarked encoder can remember these special samples, which is robust to the input noising attacks.

\subsubsection{Output Perturbing.}

The adversary can also add some perturbations to the embeddings before returning them as the outputs.
Here we consider two kinds of perturbations, i.e., random noising and truncation.

\mypara{Output Noising}
The adversary may return the perturbed embeddings by adding i.i.d.Gaussian noise as $h' = h + \epsilon_2 \cdot \mathcal{N}(0,1),$ where $h$ is the original embedding, $h'$ is the perturbed embedding, and $\epsilon_2$ is a hyper-parameter to control the noise level.
Then, we evaluate DA and WR on different $\epsilon_2$.
From \autoref{fig:robust_da_b}, we observe that DA decreases when $\epsilon_2$ increases.
For instance, when $\epsilon_2$ increases from 0.05 to 0.15, DA on STL-10 drops from 0.940 to 0.905.
However, the WR remains above 0.50 for all watermarked encoders (see \autoref{fig:robust_wr_b}), which means when we feed the embeddings with noise into the decoders, the secret vector can still be successfully extracted.
Therefore, the adversaries cannot remove the watermark even if they add random noise to the embeddings at the expense of decreasing the model's performance.

\mypara{Truncation}
The adversary may decrease the precision of the embeddings by leveraging truncation.
For instance, the adversary retains $k$ decimal places for each value in the embeddings, e.g., if $k=3$, the adversary modifies the value $1.2368$ to $1.236$, and changes $1.2368$ to $1$ when $k=0$.
\autoref{fig:robust_wr_c} and \autoref{fig:robust_da_c} shows WR and DA under different $k$.
We observe that DA has a sharp drop when $k$ decreases from 1 to 0.
Meanwhile, WR are all above 0.5 instead of MoCo, i.e., WR of MoCo drops to 0.00 when $k=0$, but the DA on STL-10 is only 0.10.
Therefore, adversaries cannot remove the watermark from the encoder while remaining its functionality.

\subsubsection{Model Modification.}

When adversaries have white-box access to the encoder, they can try to remove the watermark by modifying the encoder's parameters.
In this section, we consider three methods of model modification: watermark overwriting, model pruning, and fine-tuning.

\mypara{Overwriting}
The adversary can also leverage \SSLGuard to inject a new watermark into an SSL encoder whether or not they know that the encoder has already been injected with a watermark.
The adversary aims to generate a new watermarked encoder $F_*'$ from $F_*$ with a different key-tuple.
We want to confirm if our original watermark can remain in $F_*'$ as well.
For each $F_*'$, we measure the DA on different downstream tasks and the WR of the original key-tuple.
The results are shown in \autoref{table:overwriting}.
We observe that although we overwrite the watermarked encoder with a new key-tuple to generate a new encoder, the original watermark is still preserved, i.e., the WR of the original watermark in the new encoder is 1.00.
This indicates that the original watermark can still be preserved even if the adversary overwrites a new watermark into the model.

\begin{table}[ht]
\centering
\caption{Overwriting.}
\label{table:overwriting}
\begin{tabular}{cl|c ccc}
\toprule
   &        & SimCLR & MoCo v2 & BYOL \\
\midrule
\multirow{4}{*}{DA} & STL-10       &  0.785   &  0.888   &   0.954  \\
                     & CIFAR-10     &  0.765   &  0.685   &   0.863\\
                     & MNIST        &  0.962   &  0.955   &   0.977\\
                     & F-MNIST      &  0.885   &  0.837   &   0.905\\
\midrule
\multirow{2}{*}{WR}  & Overwriting key &  1.00  & 1.00  & 0.98\\
                     & Original key   &  {\bf 1.00}  & {\bf 1.00}  & {\bf 1.00} \\
\toprule
\end{tabular}
\end{table}

\begin{figure*}[t]
\centering
\begin{subfigure}{0.49\columnwidth}
\includegraphics[width=\columnwidth]{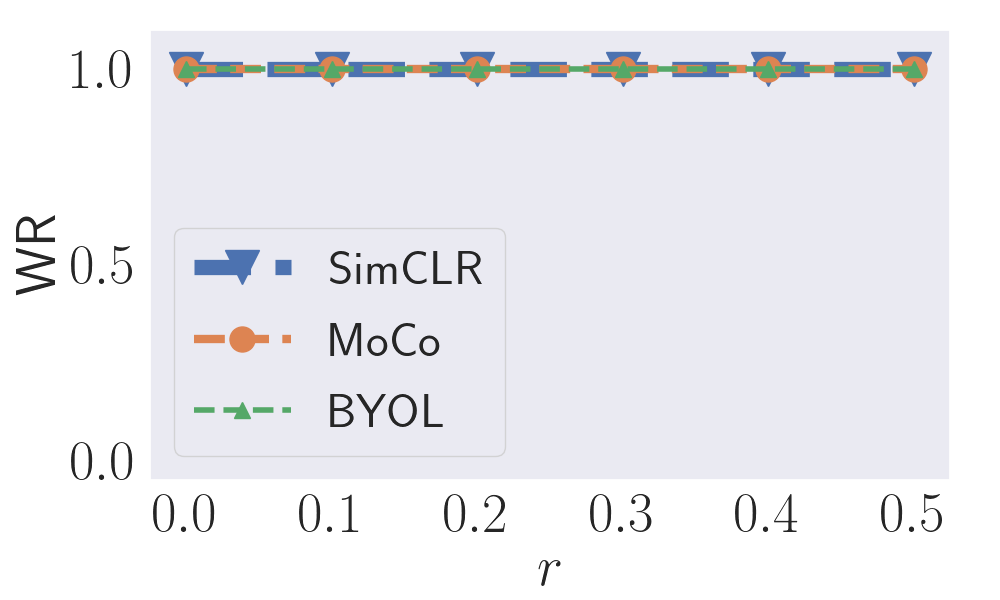}
\caption{Global pruning}
\label{fig:ft_wr_a}
\end{subfigure}
\begin{subfigure}{0.49\columnwidth}
\includegraphics[width=\columnwidth]{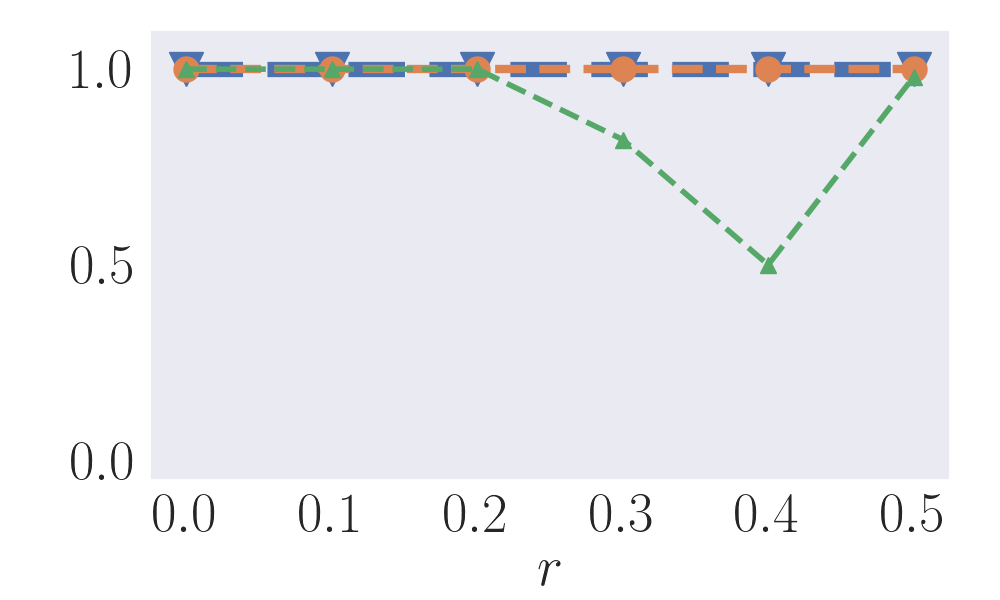}
\caption{Local pruning}
\label{fig:ft_wr_b}
\end{subfigure}
\begin{subfigure}{0.49\columnwidth}
\includegraphics[width=\columnwidth]{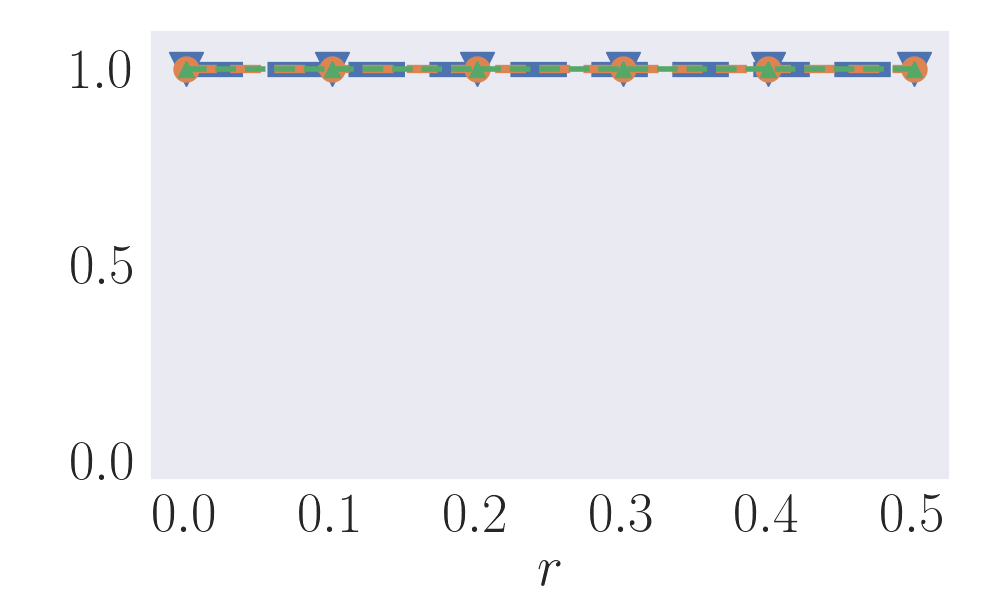}
\caption{Global fine-tuning}
\label{fig:ft_wr_c}
\end{subfigure}
\begin{subfigure}{0.49\columnwidth}
\includegraphics[width=\columnwidth]{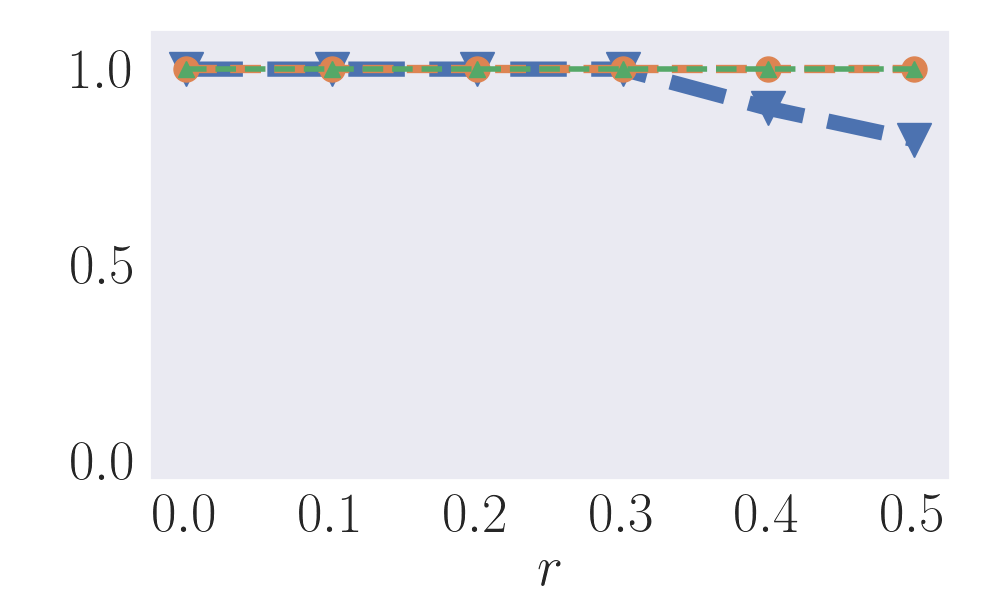}
\caption{Local fine-tuning}
\label{fig:ft_wr_d}
\end{subfigure}
\caption{The WR of pruned and fine-tuned encoders.}
\label{fig:ft_wr}
\end{figure*}

\begin{figure*}[t]
\centering
\begin{subfigure}{0.49\columnwidth}
\includegraphics[width=\columnwidth]{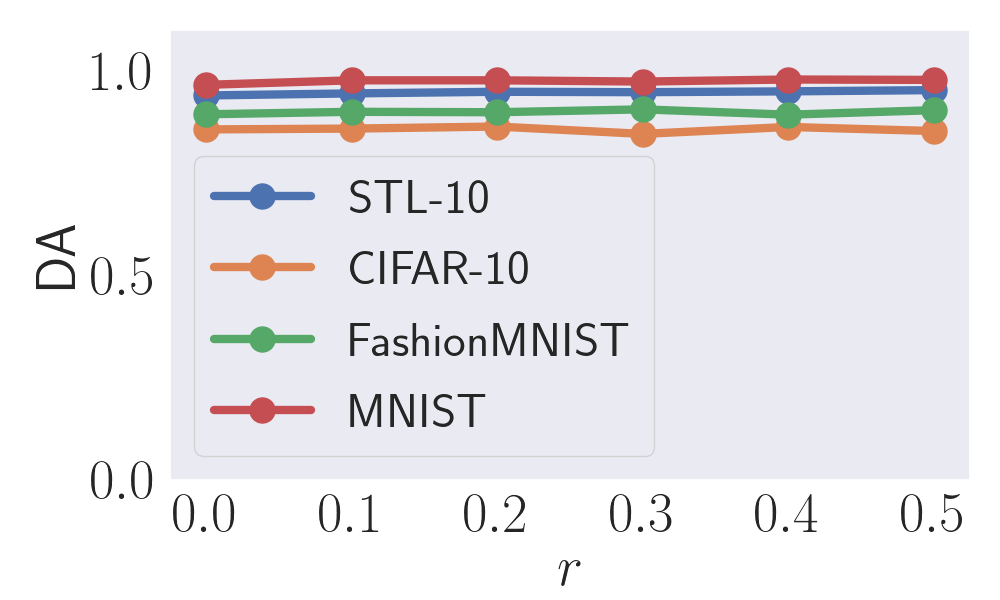}
\caption{Global pruning}
\label{fig:ft_da_a}
\end{subfigure}
\begin{subfigure}{0.49\columnwidth}
\includegraphics[width=\columnwidth]{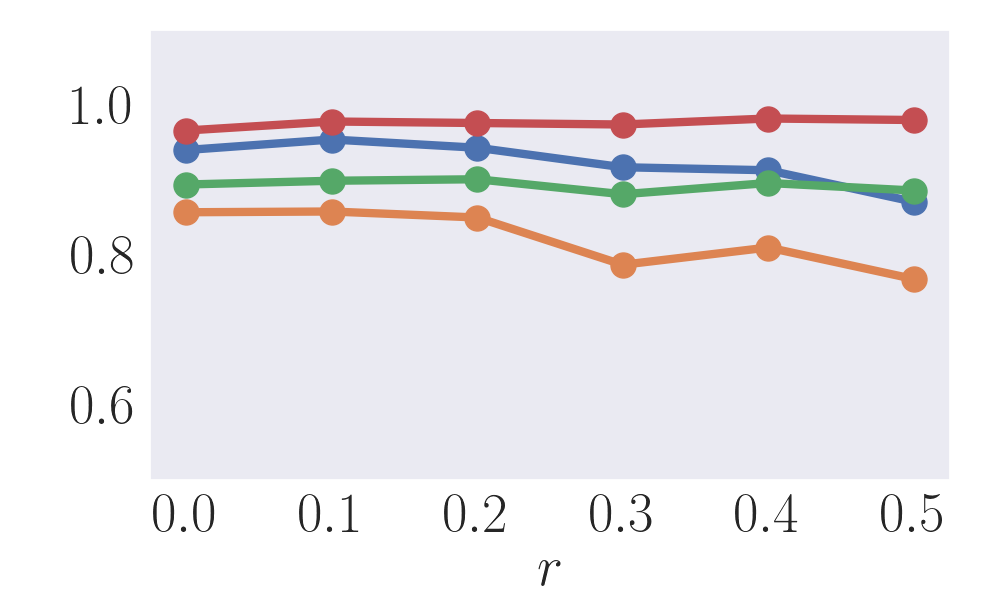}
\caption{Local pruning}
\label{fig:ft_da_b}
\end{subfigure}
\begin{subfigure}{0.49\columnwidth}
\includegraphics[width=\columnwidth]{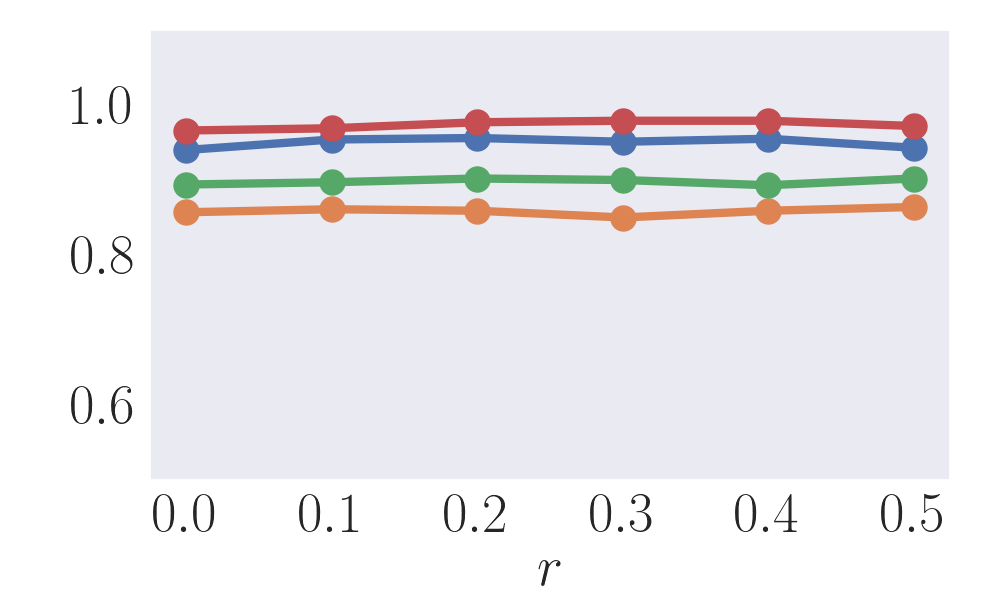}
\caption{Global fine-tuning}
\label{fig:ft_da_c}
\end{subfigure}
\begin{subfigure}{0.49\columnwidth}
\includegraphics[width=\columnwidth]{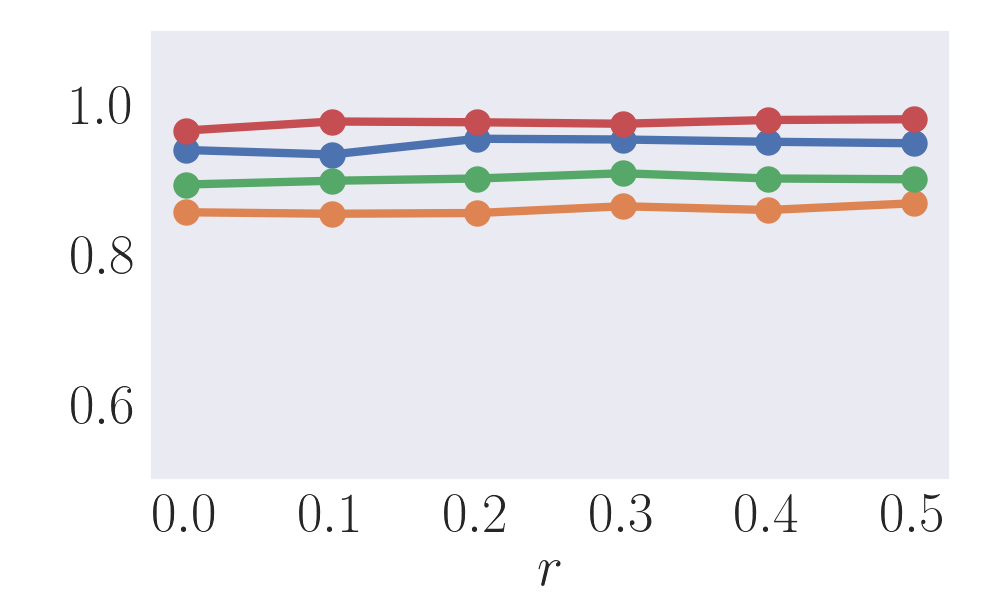}
\caption{Local fine-tuning}
\label{fig:ft_da_d}
\end{subfigure}
\caption{The DA of pruned and fine-tuned encoders. The victim encoder is BYOL.}
\label{fig:ft_da}
\end{figure*}

\mypara{Pruning}
Pruning is an effective technology for model compression~\cite{ZG18}.
It is also considered a watermark removal attack since many neurons may be disabled which reduces the effectiveness of the watermark~\cite{LJLK22}.
In this part, we leverage global and local unstructured pruning methods to the watermarked encoders.
In the global pruning setting, we set $r$ fraction of weights in the convolutional layers which have the smallest absolute values in all layers to 0.
Compared to global pruning, i.e., putting together all the connections across different layers and comparing them, local pruning aims to prune a proportion of connections with the smallest absolute values in the same layer.
We show the WR and DA in the first two sub-figures of \autoref{fig:ft_wr} and \autoref{fig:ft_da}, respectively.
We observe that DA and WR drop a little as the ratio increases in global pruning.
However, for local pruning, there is a larger downward trend in DA.
For instance, DA is 0.954 when $r=0.1$ and 0.871 when $r=0.5$, this is because local pruning cannot preserve the global information in the model properly.
In general, most of the WR are 1.0, which means \SSLGuard is robust to different pruning settings.
We also notice a special case here, i.e., on BYOL, when $r=0.4$, the WR is 0.50.
This is the worst case in our experiment, which demonstrates that we use watermark verification threshold $th_w=0.5$ in \SSLGuard is reasonable.
Also, note that for all clean encoders we evaluate in this paper, the WR is 0.
This means the $th_w$ can be set to a smaller value to better verify the watermarked encoder as we discussed in \autoref{subsection:sslguard_overview}.

\mypara{Fine-tuning}
After pruning, the adversary can fine-tune the surrogate encoders under the victim encoder's supervision, which is following the setting in~\cite{JCCP21}.
This process is also called fine-pruning~\cite{LDG18}.
The goal of fine-tuning is to regain DA's drop.
We fine-tune all the weights of the pruned encoders (global and local) by the MSE loss function.
We note that we freeze the BatchNorm layers of the pruned encoders due to reducing inaccurate batch statistics estimation caused by a small batchsize~\cite{WH18}.
The WR are shown in \autoref{fig:ft_wr_c} and \autoref{fig:ft_wr_d}, and the DA are shown in \autoref{fig:ft_da_c} and \autoref{fig:ft_da_d}.
We observe that fine-tuning can recover lost information from the victim encoder.
For instance, when $r=0.3$ in the local pruned model, DA on STL-10 is 0.917.
After fine-tuning the pruned model, DA comes to 0.954.
Meanwhile, WR increases as DA recovers.
This means \SSLGuard is robust to fine-tuning.

\subsubsection{Model Stealing.}

We then quantify the robustness of \SSLGuard through the lens of model stealing attacks.
Note that we only consider the most powerful surrogate encoder's architectures and most effective query datasets.
Concretely, based on the evaluation in \autoref{subsection:clean_pretrained_encoder}, we consider ResNet-50 and ResNet-101 as the surrogate encoder's architectures and STL-10 as the query dataset.
We name the three attacks \stealone, \stealtwo, and \stealthree.
The details of each attack are shown in \autoref{table:steal_type}.

\begin{table}[ht]
\centering
\caption{Details of different model stealing attacks.}
\label{table:steal_type}
\begin{tabular}{l| ccc }
\toprule
Attacks        & Query dataset & Architecture & Loss function \\
\midrule
\stealone      & STL-10  & ResNet-50    & Cosine \\
\stealtwo    & STL-10  & ResNet-101  & Cosine \\
\stealthree     & STL-10 (s) & ResNet-50 & Cosine \\
\bottomrule
\end{tabular}
\end{table}

The WR and DA for different attacks are shown in \autoref{table:steal_results}.
We observe that although the model stealing attack is effective against the watermarked encoder, we can still verify the ownership of the surrogate model as the WR is also high.
For instance, for \stealtwo against the watermarked encoder pre-trained by BYOL, the DA is 0.937 and 0.815 on STL-10 and CIFAR-10, while the WR is 1.00, which indicates that the watermark injected by \SSLGuard can still preserve in the surrogate encoder stolen by the adversary.
We also have similar observations on \stealone and \stealthree, which demonstrate the robustness of \SSLGuard under model stealing attacks.

\begin{table}[ht]
\centering
\caption{The DA and WR of model stealing attacks against the watermarked encoders.}
\label{table:steal_results}
\setlength{\tabcolsep}{1.1mm}{
\begin{tabular}{c|ll|ccc }
\toprule
  Attacks     & \multicolumn{2}{c|}{Metric} &SimCLR & MoCo & BYOL \\
\midrule
\multirow{5}{*}{ \stealone}  & \multirow{4}{*}{ DA}  & STL-10   & 0.721 & 0.890 & 0.938\\
                                & & CIFAR-10 & 0.685 & 0.628 & 0.791\\
                                & & F-MNIST  & 0.832 & 0.809 & 0.830\\
                                & & MNIST    & 0.928 & 0.923 & 0.915\\ \cmidrule{2-6}
                                & \multicolumn{2}{c|}{WR}  & {\bf 1.00}  &  {\bf 0.96} & {\bf 1.00}  \\
\midrule
\multirow{5}{*}{ \stealtwo}  & \multirow{4}{*}{ DA}    & STL-10   & 0.727 & 0.871 & 0.937\\
                                & & CIFAR-10 & 0.677 & 0.628 & 0.815\\
                                & & F-MNIST  & 0.840 & 0.827 & 0.865\\
                                & & MNIST    & 0.935 & 0.919 & 0.961 \\ \cmidrule{2-6}
                                & \multicolumn{2}{c|}{WR}        & {\bf 0.99}  &  {\bf 0.90}   & {\bf 1.00}  \\
\midrule
\multirow{5 }{*}{ \stealthree} &  \multirow{4}{*}{ DA}   & STL-10  & 0.732 & 0.874 & 0.923\\
                                & & CIFAR-10 & 0.677  & 0.658 & 0.784\\
                                & & F-MNIST  & 0.827  & 0.823 & 0.851\\
                                & & MNIST   & 0.932  & 0.940 & 0.922\\ \cmidrule{2-6}
                                & \multicolumn{2}{c|}{WR}       & {\bf 1.00}  &  {\bf 0.95}  & {\bf 0.98}  \\
\bottomrule
\end{tabular}
}
\end{table}

\section{Discussion}

\mypara{The Necessity of the Shadow Encoder}
The reason why \SSLGuard can extract watermarks from the surrogate encoder is that it locally simulates a model stealing process by using a shadow dataset and shadow encoder.
In this part, we aim to demonstrate the need for such a design.
We discard the shadow encoder and inject the watermark into a clean pre-trained encoder on SimCLR, MoCo v2, and BYOL.
Then we get the corresponding key-tuples.
The key-tuples can extract watermarks successfully.
However, when We mount \stealone to the watermarked encoders to generate three surrogate encoders (i.e., $S^{simclr}$, $S^{moco}$, and $S^{byol}$), the WR are all 0.00, which means the watermark may not be verified.
Meanwhile, DA for $S^{byol}$ are 0.945, 0.735, 0.843, and 0.926 on STL-10, CIFAR-10, F-MNIST, and MNIST, respectively.
This indicates that the adversary can successfully steal the victim encoder as the DA for the surrogate encoder are close to the target encoder.
In conclusion, \SSLGuard cannot work well without the shadow encoder as the adversary can steal a surrogate encoder with high utility while bypassing the watermark verification process.
Therefore, the shadow encoder is crucial for defending against model stealing attacks.

\mypara{The Choice of Mask}
In our experiments, we set the covering space of the mask as $35\%$.
We also leverage different masks $M$, i.e., $5\%$ and $50\%$ to inject watermark into BYOL, then we mount \stealone to the watermarked encoders, the WR are 0.99 and 1.00.
The results show that the WR is similar when we leverage different covering spaces of the masks, which indicates that \SSLGuard is effective under different masks.

\mypara{Extension to Other Types of Datasets}
In this paper, we only focus on encoders pre-trained on image datasets.
To extend \SSLGuard into encoders pre-trained on other types of datasets such as texts or graphs~\cite{GNWB21,YCSCWS20}, the main challenge is to define a suitable trigger pattern in the language or graph domain.
Then we can apply a similar method to watermark those models.
We leave it as our future work to further explore the effectiveness of \SSLGuard on other domains such as texts or graphs.

\section{Related Work}

\mypara{Privacy and Security for SSL}
There have been more and more studies on the privacy and security of self-supervised learning.
Jia et al.~\cite{JLG21} sum up 10 security and privacy problems for SSL.
Among them, only a small part has been studied.
Liu et al.~\cite{LJQG21} study MIA against contrastive learning-based pre-train encoder.
Concretely, Liu et al.~\cite{LJQG21} leverage data augmentations over the original samples to generate multiple augmented views.
Then, the authors measure the similarities among the embeddings of the augmented samples.
The intuition is that, if the sample is a member, then the similarities should be higher than a non-member.
He and Zhang~\cite{HZ21} perform the first privacy analysis of contrastive learning.
Concretely, the authors observe that the contrastive models are less vulnerable to membership inference attacks, while more vulnerable to attribute inference attacks.
The reason is that contrastive models are more generalized with less overfitting level, which leads to fewer membership inference risks, but the representations learned by contrastive learning are more informative, thus leaking more attribute information.
Jia et al.~\cite{JLG22} propose the first backdoor attack against SSL pre-trained encoders.
By injecting the trigger pattern in the pre-training process of an encoder that correlated to a specific downstream task, the backdoored encoder can behave abnormally for this downstream task.
The author further shows that triggers for multiple tasks can be simultaneously injected into the encoder.

\mypara{DNNs Copyright Protection}
In recent years, several techniques for DNNs copyright protection have been proposed.
Among them, DNNs watermarking is one of the most representative algorithms.
Jia et al.~\cite{JCCP21} propose an entangled watermarking algorithm that encourages the classifiers to represent training data and watermarks similarly.
The goal of the entanglement is to force the adversary to learn the knowledge of the watermarks when he steals the model.
DNN fingerprinting is another protection method.
Unlike watermarking, the goal of fingerprinting is to extract a specific property from the model.
Cao et al.~\cite{CJG21} introduce a fingerprinting extraction algorithm, namely IPGuard.
IPGuard regards the data points near the classification boundary as the model's fingerprint.
If a suspect classifier predicts the same labels for these points, then it will be judged as a surrogate classifier.
Chen et al.~\cite{CWPSCJMLS22} propose a testing framework for supervised learning models.
They propose six metrics to measure whether a suspect model is a copy of the victim model.
Among these metrics, four of them need white-box access, and black-box access is enough for the rest.

\section{Conclusion}

In this paper, we first quantify the copyright breaching threats of SSL pre-trained encoders through the lens of model stealing attacks.
We empirically show that the SSL pre-trained encoders are highly vulnerable to model stealing attacks.
This is because the rich information in the embeddings can be leveraged to better capture the behavior of the victim encoder.
To protect the copyright of the SSL pre-trained encoder, we propose \SSLGuard, a robust black-box watermarking scheme for the SSL pre-trained encoders. 
Concretely, given a secret vector, \SSLGuard injects a watermark into a clean pre-trained encoder and outputs a watermarked version.
The shadow training technique is also applied to preserve the watermark under potential model stealing attacks.
Extensive evaluations show that \SSLGuard is effective in embedding and extracting watermarks and robust against model stealing and different types of watermark removal attacks such as input noising, output perturbing, overwriting, model pruning, and fine-tuning.

\section*{Acknowledgement}

We thank all anonymous reviewers for their constructive comments.
This work is partially funded by the Helmholtz Association within the project ``Trustworthy Federated Data Analytics'' (TFDA) (funding number ZT-I-OO1 4), by the National Key Research and Development Program of China (2018YFA0704701, 2020YFA0309705), by the Major Program of Guangdong Basic and Applied Research (2019B030302008), and by the Major Scientific and Technological Innovation Project of Shandong Province (2019JZZY010133).

\newpage
\bibliographystyle{plain}
\bibliography{normal_generated_py3}

\end{document}